# Induced Giant Piezoelectricity in Centrosymmetric Oxides

D.-S. Park[1,2*], M. Hadad[2], L. M. Riemer[1], R. Ignatans[3], D. Spirito[4], V. Esposito[5], V. Tileli[3], N. Gauquelin[6], D. Chezganov[6], D. Jannis[6], J. Verbeeck[6], S. Gorfman[4], N. Pryds[5], P. Muralt[2], D. Damjanovic[1*]

[1]Group for Ferroelectrics and Functional Oxides, Swiss Federal Institute of Technology–EPFL, 1015 Lausanne, Switzerland.

[2]Group for Electroceramic Thin Films, Swiss Federal Institute of Technology–EPFL, 1015 Lausanne, Switzerland.

[3]Institute of Materials, Swiss Federal Institute of Technology–EPFL, 1015 Lausanne, Switzerland.

[4]The Department of Materials Science and Engineering, Tel Aviv University, Ramat Aviv, Tel Aviv 6997801, Israel.

[5]Department of Energy Conversion and Storage, Technical University of Denmark, Fysikvej, 2800 Kgs. Lyngby, Denmark.

[6]Electron Microscopy for Materials Science (EMAT), University of Antwerp, B-2020 Antwerpen, Belgium.

*E-mail: dspark1980@gmail.com, dragan.damjanovic@epfl.ch

**Abstract:**

**Piezoelectrics are materials that linearly deform in response to an applied electric field. As a fundamental prerequisite, piezoelectric material must possess a non-centrosymmetric crystal structure. For more than a century, this remains a major obstacle for finding piezoelectric materials. We circumvented this limitation by breaking the crystallographic symmetry, and inducing large and sustainable piezoelectric effects in centrosymmetric materials by electric field-induced rearrangement of oxygen vacancies. Our results show the generation of extraordinarily large piezoelectric responses ($d_{33}$ ~200,000 pm/V at mHz frequencies), in cubic fluorite Gd-doped $CeO_{2-x}$ films, which is two orders of magnitude larger than in the presently best-known lead-based piezoelectric relaxor-ferroelectric oxide at kHz frequencies. These findings open opportunities to design piezoelectric materials from environmentally friendly centrosymmetric ones.**

The fundamental principle of electrostriction and piezoelectric effects stems from small deformations of the crystal unit cell by an applied electric field. The latter effect also provides charge separation under mechanical pressure (1). The associated displacements of atoms are in the picometer range so that atoms remain confined around their original crystallographic sites. Piezoelectricity is of a high technological and industrial importance and is employed in a vast number of applications such as medical devices, actuators, and sensors (2). Motivations to augment the piezoelectric response, which requires materials with a non-centrosymmetric structure, are thus compelling. Among the piezoelectric materials, perovskite-type oxides are the most widely used and exhibit excellent piezoelectric responses. Several routes to achieve the highest electromechanical response in these materials have been pursued and these include control of the material's structural instability at specific chemical compositions (e.g. morphotropic phase boundary) and associated polarization rotation and domain engineering (3), chemical disorder (4), and nanocomposite structures (5). All of these strategies demonstrate the possibility for improving the piezoelectric response within an order of magnitude with respect to that of the industrial standard, $Pb(Zr,Ti)O_3$ (PZT) (6).

The piezoelectric effect can be induced in centrosymmetric materials by applying a direct electric field which breaks the inversion symmetry (7, 8). This approach has been revived recently by applying asymmetric electrodes on centrosymmetric samples creating different Schottky barriers at the electrodes (9). This approach has potential to widen the number of prospective electromechanical materials beyond the traditionally dominating ferroelectric lead-based perovskites, but the resulting response is still one to two orders of magnitude lower than that of PZT. Another attempt to induce the effect suggests using electric field-assisted exchange of oxygen with the ambient atmosphere in oxygen-nonstoichiometric fluorite structures (Y-doped $ZrO_2$, $Re^{+3}$-doped $CeO_2$, where Re is a rare earth element) (10). The oxygen exchange (release) results in chemical expansion of the films, leading to bending of thin materials and thus the strain–electrical field ($x$ vs $E$) relationship which mimics the piezoelectric effect. The strains achieved are comparable to those in PZT. The field-induced motion of charge carriers take place over a long-range (size of the sample), and the effect is large only at high temperatures (> 500°C) and low frequencies (below ~1 Hz), where ionic diffusion is sufficiently large. In addition, some fluorite oxides (*e.g.*, doped $CeO_2$) were reported to exhibit electrostrictive coefficients that are about two orders of magnitude higher than expected values from phenomenological relations between electrostrictive coefficient and elastic and dielectric susceptibilities (11). The origin of this large electrostriction has not been entirely yet elucidated, but the electric field-induced mechanical deformation is certainly related to the presence and short-range motion of oxygen vacancies (*12-14*).

We demonstrate a paradigm shift for achieving large, electric field-induced piezoelectricity in centrosymmetric materials. We show that for Gd-doped CeO$_{2-x}$ (CGO) films, which possess a cubic fluorite centrosymmetric structure, we can achieve very high values of the electric-field-induced piezoelectric strain (*x* ~26%) and apparent longitudinal piezoelectric coefficients (*d$_{33}$* ~200,000 pm/V). Only for purpose of illustrating the size of the achieved coefficients, we note that the latter value, measured in the mHz range, is two to three orders of magnitude larger than that in the best piezoelectric perovskite oxides, e.g. Pb(Mg$_{1/3}$Nb$_{2/3}$)O$_3$-PbTiO$_3$ with *d$_{33}$* ≈ 2000 pm/V [3]. Importantly, and relevant for applications, the induced effect is comparable to that in the best PZT thin films (~100 pm/V) in the kHz range (*15*). We argue that the change in the strain mechanism from the short-range lattice/ionic defect-based mechanism above 10 Hz to the one at low frequencies, is based on distinct actions of long-range migration of ions (oxygen vacancies, V$_O$) and electrons. Our results show that the electric field-induced redistribution of mobile charges in the films leads to crystal phase transition associated with chemical expansion, and material heterogeneity. These combined effects result in giant piezoelectric and electrostrictive responses and points towards previously unknown electromechanical mechanism in centrosymmetric fluorites and materials with large ionic and electronic conductivity in general.

We deposited polycrystalline (Gd$_{0.2}$Ce$_{0.8}$)O$_{2-x}$ films on Al/SiO$_{2-x}$/Si(100) substrate at room temperature by sputter deposition (Fig. 1A). The CGO films had thicknesses in the range of ~1.25 - ~1.8 µm (Fig. S1) (*16*). The electrostrictive strain for a sample of length, *L*, is defined as:

$$x = \Delta L/L = ME^2, \qquad (eq.1)$$

where *x* is the out of plane strain of the film, $\Delta L$ is the thickness change, *M* the corresponding electrostrictive coefficient, and *E* the alternated electric field applied between films top and bottom electrodes, $E = E_{AC}(t) = E_{AC}\sin(2\pi ft)$, where *f* is the frequency. We employed two different top electrodes, Al (~150 nm) and Pt (~150 nm)/Cr (~20 nm) layers to study the effect of asymmetric electrodes on the polarization (Fig. S2C). Electrical (current density, *J vs* electric field, *E*) and electromechanical (strain, *x* vs *E*) measurements on the films were performed on *out-of-plane* capacitor geometry. We measured the strain, *x*, using a contactless fiber-optic method (*16*). We confirmed the electromechanical displacements using another setup with direct contact measurements (Section 1). All of the instrumental artefacts and external effects (*e.g.*, bending and Joule heating effects) have negligible influence and thus the dominant contribution to the induced strain is originated only from the electromechanical response of the films (Figs. S2,S3).

We show the measured $\Delta L$ with the second harmonic response and an offset as obtained from $[\sin(2\pi ft)]^2 = \frac{1}{2}(1 - \cos(4\pi ft))$ (Fig. 1B). We measured the electric current density, $J$, through the film simultaneously with the change in the length, $\Delta L$, and charge density, $D = \int Jdt$ (Fig. 1B, section 2, fig. S4B) (*16*). We performed measurements of electrostriction across a range of frequencies from 3 mHz to 1 kHz. The electrostrictive coefficient, $M$, which we determined using eq. 1 shows a remarkably complex frequency dependence (Fig. 1C). This clearly indicates that at least three different contributions to the electrostriction exist in these samples (Fig. 1C, fig. S4). We suggest that the strong rate-dependent contributions to strain can be attributed to the existence of mobile ionic species ($V_O$) in the CGO film, as implied by a similar behavior in the AC conductivity (Figs. S4,S5).

The inversion symmetry in CGO can be broken by applying an electric-field-bias, $E_{DC}$, leading to asymmetric charge distribution and induced polarization, $P_{ind}$, in the material (Fig. 2A) (*1*). We can explain this by replacing field, $E$, in eq. 1 by $E = E(t)_{AC} + E_{DC}$:

$$x = ME_{DC}^2 + ME_{AC}^2 + (2ME_{DC})E_{AC}. \quad (\text{eq.2})$$

The first and second term describe electrostrictive deformations, while the third term is the symmetry-breaking term with the field-induced piezoelectric coefficient, $d_{ind} = 2ME_{DC}$.

The CGO possesses a centrosymmetric cubic fluorite structure in the ground state, and is not piezoelectric. However, we observed the piezoelectric displacement term ($d_{ind} = 2ME_{DC}$) in our CGO films upon application of electric field bias (Fig. 2B) for $E_{DC} = \pm 0.47$ MV/cm and $E_{AC} = 15.71$ kV/cm. We observe a clear presence of the first harmonic deformation as well as a 180° phase shift when changing the sign of $E_{DC}$, which correspond to the induced piezoelectric effect. We do not observe in this case the electrostrictive displacements (2nd harmonic) (Figs. 2B,C) due to a very small amplitude of $E_{AC}$ (compared to $E_{AC}$ in Fig. 1A). Once the $E_{AC}$ is comparable to or higher than $E_{DC}$, we observed an asymmetric response, comprising both the first and the second harmonics (Fig. S6). Measuring the piezoelectric strain of the CGO sample as a function of $E_{DC}$ (in the range of ±0.47 MV/cm), while keeping the same electric field $E_{AC}$ (15.71 kV/cm) at 10 mHz, clearly shows that the applied $E_{DC}$ tunes the piezoelectric AC strain, reaching values of up to 2.15 % ($d_{ind}$ ~13,700 pm/V) (Fig. 2D). The nonlinear dependence of the AC strain on $E_{DC}$ (Fig. 2D) reflects the fundamental nature of the electrostriction coupling to polarization, and not directly to the field. Furthermore, the on-off control of piezoelectricity and electrostriction, and the tuning of the electromechanical response can be sustained for at least several hours without any sign of degradation (Figs S7,S8).

We show the piezoelectric coefficients of the CGO sample, determined as a function of frequency (from 10 mHz to 1 kHz) for different fields, $E_{DC}$, *e.g.*, 0.47, 0.72, and 1.00 MV/cm (Fig. 3A). The results are remarkable when compared with the frequency-independent response in conventional piezoelectric materials, *e.g.*, PZT and a bismuth titanate-based ceramic (Fig. S9). First, the piezoelectric coefficient reaches giant values at low frequencies, approaching ~200,000 pm/V with increasing $E_{DC}$. For comparison, the piezoelectric coefficient in the best commercial single crystals of Pb(Mg$_{1/3}$Nb$_{2/3}$)O$_3$-PbTiO$_3$ (PMN-PT) is ~2,000 pm/V, and in PZT ceramics ~200 - 500 pm/V [3]. More importantly, the values around 100 pm/V, measured at 1 kHz in our films are comparable to those of PZT thin films (*15*). This is the frequency range of interest for many actuator applications. We observed clear high-*f* (1 kHz) piezoelectric responses for the CGO film with a linear relation following $x = d_{ind}E_{AC}$, while, as expected, the $d_{ind}$ varies with applied $E_{DC}$ (Fig. 3B).

Further insight into the electric-field-induced piezoelectric response of the CGO can be obtained from the piezoelectric term of Eq. (2) from which one can derive:

$$d_{ind} = 2ME_{DC} = 2\varepsilon Q P_{ind}. \quad (eq.3)$$

where $\varepsilon$ is the dielectric permittivity, $Q$ the polarization-related electrostrictive constant, $x = QP_{ind}^2$ (*17*), $P_{ind} = \varepsilon E_{DC}$ is the induced polarization, and $M = \varepsilon^2 Q$. Eq. 3 holds in general very well for centrosymmetric materials, for example, for perovskite relaxors (*7*) and Schottky barrier-induced piezoelectric effect [9]. We show the simultaneously measured piezoelectric and electrostrictive coefficients over a wide frequency range (Fig. 3C). The ratio of $|d|/|M|$ is expected from eq. 3 to be equal to $2E_{DC}$. We see a good agreement ($|d|/|M| \approx 2E_{DC}$) for $f \geq 10$ Hz (Fig. 3D), while the ratio is lower at low frequencies ($f \leq 1$ Hz). Consequently, these data indicate presence of a rate-dependent mechanism that is triggered by the application of $E_{DC}$, and which is assisted by application of quasi-static $E_{AC}$ at low frequencies. The relationship, $d_{ind} = 2\varepsilon Q P_{ind}$, is considered to be fundamental and always holds (*1, 17*), where the polarization response is controlled by small oscillations of ions and electrons near their equilibrium lattice sites. Our results show that good agreement between the calculated and the measured $M$ and $d$ values holds over the frequency range where apparent polarization and permittivity ($\varepsilon_{ij} = \frac{\partial P_i}{\partial E_j}$) are dominated by the rate-dependent migration of $V_O^{\bullet\bullet}$ (Section 3,figs. S10,S11) (*16*). The introduction of aliovalent dopants, *e.g.*, $Gd^{3+}$ in $CeO_2$, produces negative charges and hence requires 1/2 oxygen vacancy for maintaining charge neutrality; in Kröger-Vink notation (*18*) this can be written as, $Ce^{\times}_{0.8-2y}Ce'_{2y}(Gd'_{Ce})_{0.2}O^{\times}_{1.9-y}(V_O^{\bullet\bullet})_{0.1+y}$, where $[Gd'_{Ce}] = 2[V_O^{\bullet\bullet}]$. The additional oxygen vacancies (y) produced during preparation and charge compensated by $Ce^{+3}$ are arguably more mobile than those associated with Gd (*14, 19*), at least at weaker fields. Importantly, both motion of $V_O^{\bullet\bullet}$ and polarons hopping from $Ce^{+3}$ to $Ce^{+4}$ have a substantial effect on local lattice strain through chemical expansion as well as on polarization (*20*). From symmetry arguments, only those defects that are simultaneously electric and elastic dipoles can contribute to the piezoelectric effect (*21*). We observed evidence for charge transport in the electrical conductivity of the CGO films typical for hopping-like ion conduction below 1 kHz. The ionic conductivity greatly increases by applying higher fields (both $E_{AC}$ and $E_{DC}$) (Fig. S4C,fig. S12). The conductivity seems to contribute to the giant apparent dielectric permittivity ($|\varepsilon_r| \sim 10^9$) of the system when $f \to 0$. Therefore, the defect migration is enhanced by the static $E_{DC}$ field and the quasi-static $E_{AC}$, and substantially contributes to the large permittivity leading to exceptionally large *M* and *d* at low frequencies. Strikingly, the approach for CGO with electric field is generally valid also for other systems with centrosymmetric fluorite structures, in films and bulk. We also show that piezoelectric response can be induced in a YSZ film, YSZ and CGO ceramics, as well as CGO films prepared by a different deposition technique (Figs. S13,S14,S15). The obtained piezoelectric coefficients of 10 – 100 pm/V in the frequency range of *f* = 1 Hz - 1 kHz are comparable to those presently used in MEMS device applications with materials based on (Al,Sc)N and PZT (*15*), indicating the strength of the proposed methodology.

In order to understand the relationship between the rearrangement of oxygen defects and the associated large low-frequency strain in the CGO films, we conducted *in-situ* X-ray diffraction (XRD) measurements under the application of different $E_{DC}$ electric fields. In these experiments, we directly observed partial transformation of initial, highly strained cubic-like CGO [$a_{(C)} = b_{(C)} = c_{(C)} = 5.61$ Å, $\alpha = \beta = \gamma = 90°$ for *Z* = 4, where *Z* is the number of formula units in unit cell] into a tetragonal phase [$a_{(T)} = b_{(T)} = 3.94$ Å, $c_{(T)} = 6.42$ Å, $\alpha = \beta = \gamma = 90°$ for *Z* = 2]. For instance, see the gradual appearance of the peak at $2\theta \sim 32°$ when the applied electric field (0 to 1 MV/cm) was gradually increased (Fig. 4A,fig. S16,section 4) (*16*). During the transformation, the base plane of the CGO unit cell shrinks by -0.73 %, while the *c*-axis expands by +14.39 %, resulting in a volume increase by +12.73 % (Section 4). These results explain and support the large positive strain observed along the electric field, applied along the crystallographic [001] direction, as well as the large compressive stress observed in the film plane. A similar electric field-induced phase transition has been reported in Y-doped $ZrO_2$ (YSZ) at 550°C (*22*). However, in our case the effect observed at room temperature and the resulting strains are much larger than the one reported in Ref. (*22*). In analogy to YSZ, the phase transition occurs because two oxygen sublattices are rearranged along the *c*-direction in that two oxygen atoms move up, and the others two move down while keeping a rotation symmetry of 2. The vertical spacings between O-sites remain at *c*/2, resulting in a screw axis $4_2$ along with *c*. This alternating shifting of O-columns leads to a large expansion of the *c*-axis. In YSZ, this mechanism was found to be triggered by a high $V_O$ concentration (*23, 24*), while in ceria it is believed to be also associated with a high Ce´ ($Ce^{3+}$) concentration, because in both cases (oxygen vacancy formation and the following change of host cation radii) the Coulombic attraction in the ionic matter is reduced (*20*).

These experimental results together with our measured dielectric data (Fig. S12) indicate that a field-driven defect redistribution in the film is accompanied by a partial phase transition from a cubic to a tetragonal phase accompanied with a large volumetric increase. These results are supported by the recent work of Zhu, *et al.* (*25*) which showed experimentally and theoretically that oxygen vacancies indeed play an essential role in stabilization of tetragonal phase in ceria. This suggests that the same mechanism is likely to occur in Gd-doped $CeO_2$, which may be even more susceptible to phase transition due to a higher level of oxygen vacancies, leading to electric-field driven tetragonal phase. The field-induced heterogeneity in the material is probably accompanied by Maxwell-Wagner dielectric and electromechanical effects (*26*), which together with phase transition and chemical expansion (*20*) contribute to the field-induced strain and polarization. Based on the above results, the emergent piezoelectric behavior in Gd-doped ceria directly depends on the rate-dependent $V_O$ motion on different scales (Fig. S17). The selection of different aliovalent dopants and co-doping, which can stabilize and control the presence of oxygen vacancies within an oxide material, could be an important strategy to generate sustainable large piezoelectricity using similar working mechanism as here.

We show the possibility of generating extraordinarily high piezoelectricity in intrinsically centrosymmetric nonstoichiometric oxides (fluorites) by electric-field induced redistribution of mobile $V_O$. Our results show giant low-frequency piezoelectricity (up to $d_{33}$ ~ 200,000 pm/V) in CGO films, induced by concurrent application of alternating and static electric fields. Furthermore, we show a direct way to achieve selective electro-mechanical conversion in centrosymmetric materials, *i.e.*, either pure and large electrostriction, pure and giant piezoelectricity or mixed response under controlled electric fields. Controlling chemical expansion, phase transitions, diffusion, and redistribution of mobile ionic species in centrosymmetric ionic materials by electric field is a phenomenological concept with aims to induce large electromechanical conversion, which can be extended to other material systems. Our findings provide a paradigm shift in piezoelectricity by utilizing centrosymmetric materials with large ionic mobility, and open up a path for a wide range of potential electromechanical environmentally friendly and biocompatible materials for applications in actuators and sensors.

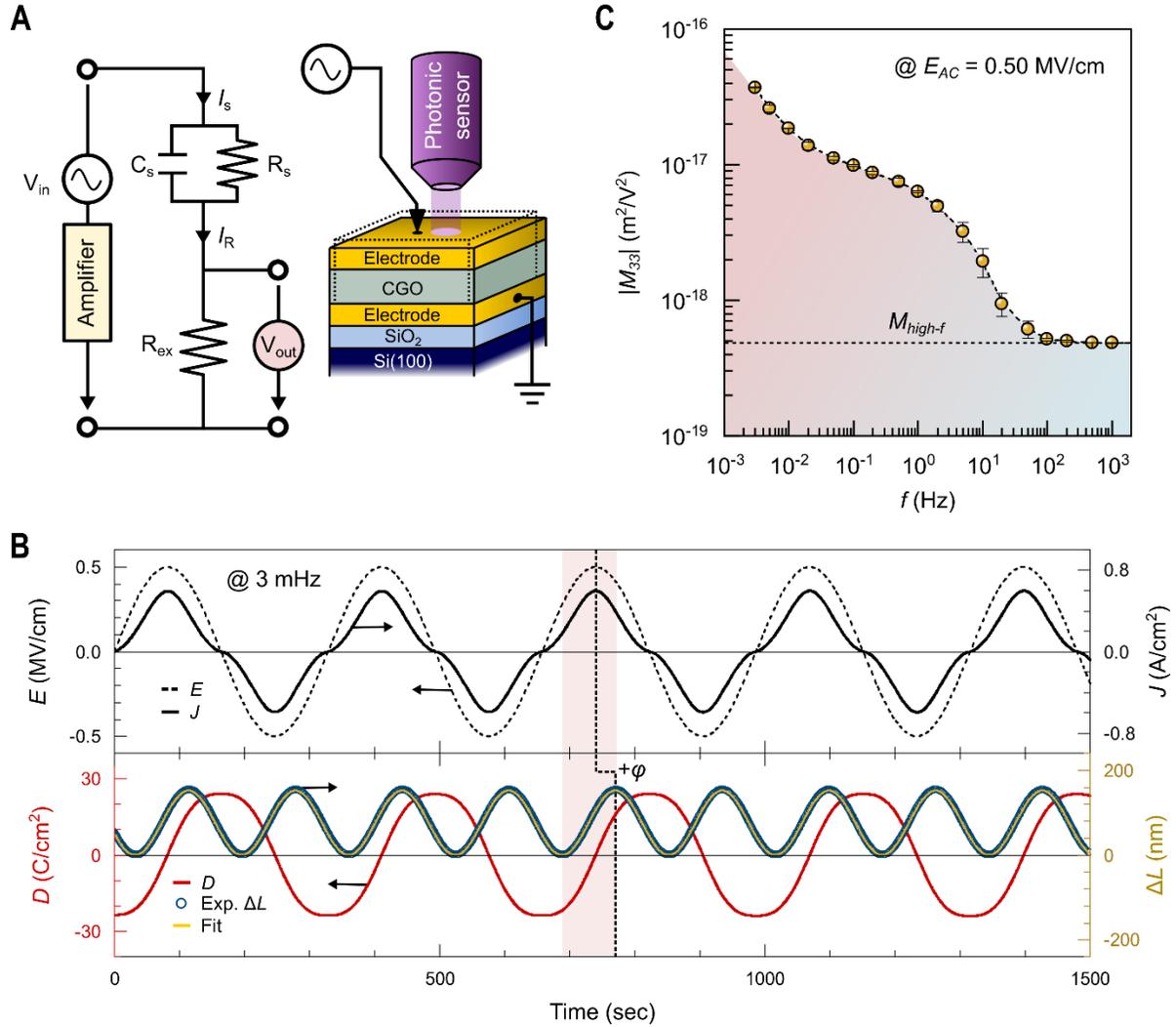

**Fig. 1. Electric field-induced electrostrictive responses of CGO film.** (**A**) Schematics of the experimental setup which combines electrical and electromechanical measurements for the CGO samples. The equivalent circuit shows the voltage source, $V_{in}$; voltage amplifier; the CGO capacitor, $C_S$, with resistance, $R_S$; an external resistor, $R_{ex}$; current, $I_R$, flowing through $R_{ex}$, and the sample; and output voltage, $V_{out}$, across $R_{ex}$. (**B**) Electrical and electromechanical outputs: (*i*) the applied $E_{AC} = 0.5$ MV/cm at $f = 3$ mHz (dashed line), (*ii*) the corresponding J (solid line), (*iii*) the derived charge density, $D$, (red solid line) and (*iv*) the concurrently measured 2nd-harmonic electromechanical response, $\Delta L$, of the samples (blue circle in lower panel). The measured $\Delta L$ in time was fitted by $\Delta L = L_0 \sin^2(\omega t + \varphi)$ as depicted by yellow solid line. (**C**) Frequency-dependent $M_{33}$ of the CGO film, excited by $E_{AC} = 0.5$ MV/cm.

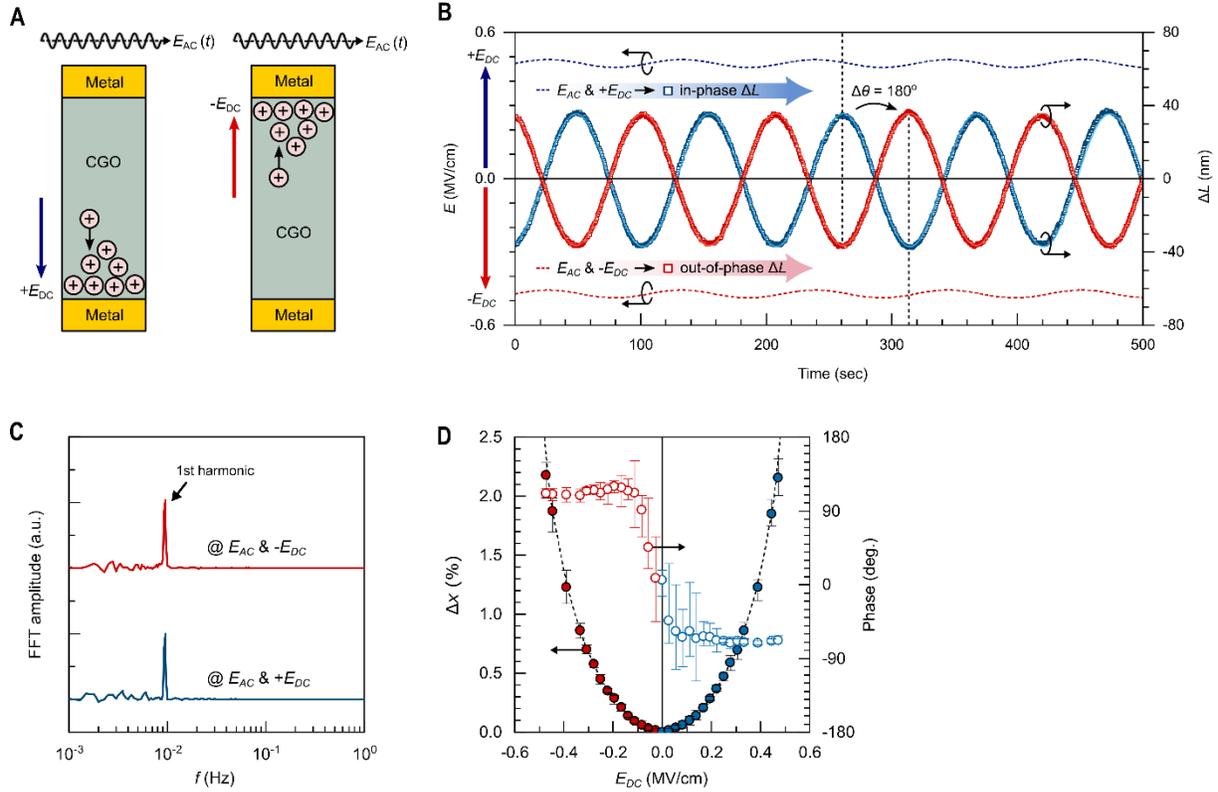

**Fig. 2. The induced piezoelectric displacements in CGO films.** (**A**) A schematic for the electric field application to CGO film in out-of-plane capacitor geometry, which is simultaneously composed of a small driving AC field ($E_{AC}$) and a large static DC field ($E_{DC}$). The induced in-phase strain ($x_{33}$) in the film is determined by the polarity of applied DC bias. (**B**) Time-resolved 1st-harmonic electro-mechanical displacements of the CGO film, measured at around 10 mHz (9.4 mHz) and excited by $E_{AC}$ =15.71 kV/cm under $E_{DC}$ = ±0.47 MV/cm. The polarity of the DC field switches the sign of piezoelectric coefficient. The measured $\Delta L$ in time was fitted by the 1st-order sine function, $\Delta L = L_0 \sin(\omega t - \varphi)$ as depicted by red and blue solid lines. (**C**) The corresponding fast Fourier transform (FFT) amplitude spectra of the generated 1st harmonic displacements in $f$. (**D**) Variations in the $x_{33}$ and response phase angle of the film as a function of DC field while applying a constant $E_{AC}$ (15.71 kV/cm).

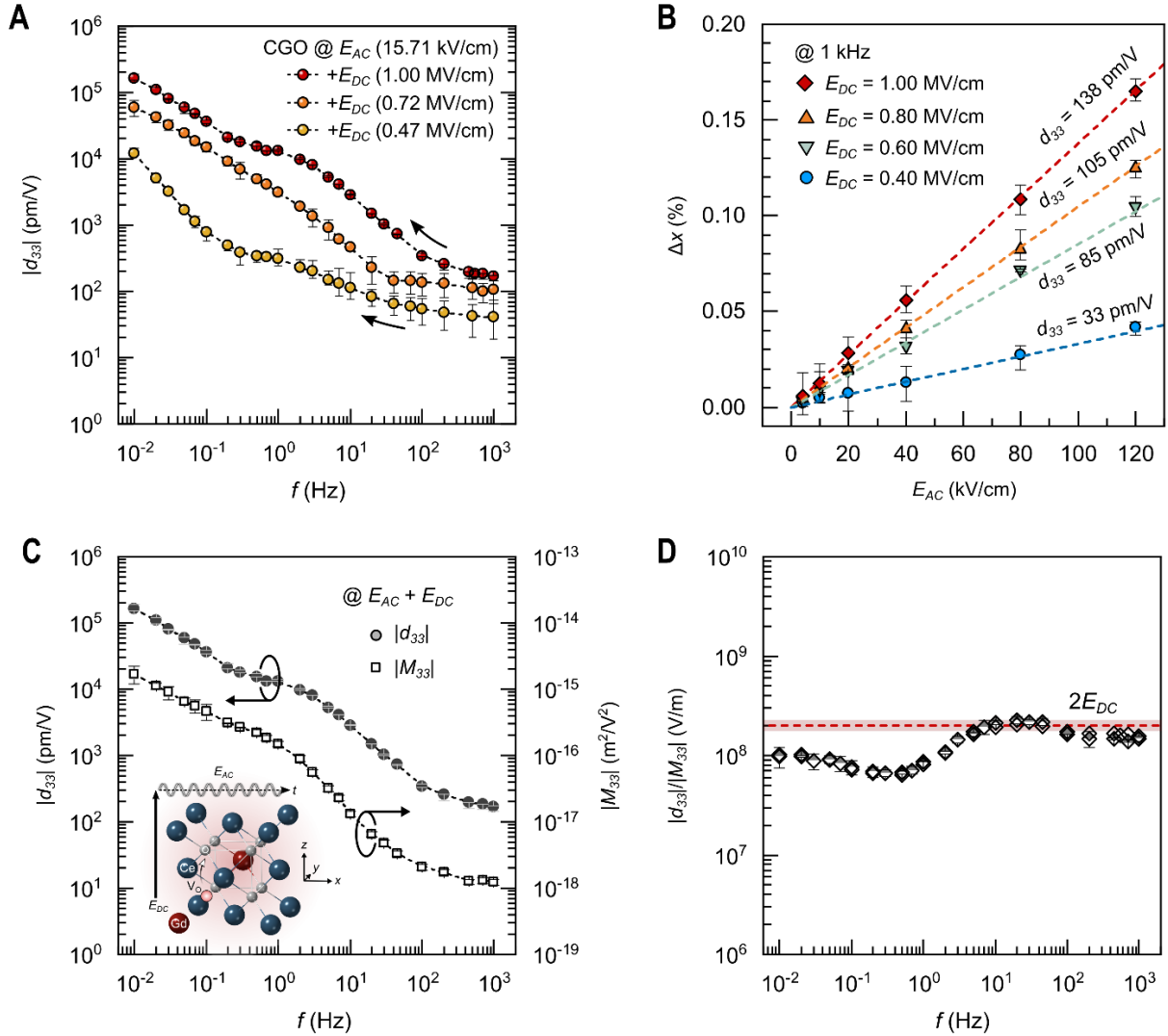

**Fig. 3. Frequency-dependent piezoelectric susceptibility of CGO film.** **(A)** The 1st-harmonic electromechanical susceptibility, $|d_{33}|$, of the CGO film as a function of $f$ (10 mHz ≤ $f$ ≤ 1 kHz), excited by a constant driving AC field ($E_{AC}$ = 15.71 kV/cm) and different static DC fields ($E_{DC}$ = +0.47, +0.72, and +1.00 MV/cm). **(B)** Linear piezoelectric strain of the CGO film as a function of $E_{AC}$ with various $E_{DC}$, measured at 1 kHz. **(C)** $|d_{33}|$ and $|M_{33}|$ of the film, simultaneously measured by applying a combined electric field, $E_{AC}$ = 15.71 kV/cm and $E_{DC}$ = +1.00 MV/cm, in the frequency range from 10 mHz to 1 kH. The inset describes the field-enforced defect dynamics, polarization reorientation, and the following permittivity variations in CGO. **(D)** Ratios of $|d_{33}|$ to $|M_{33}|$ as a function of $f$, expected to be $2E_{DC}$ in the relation of $d_{ind} = 2M_{ind}E_{DC}$.

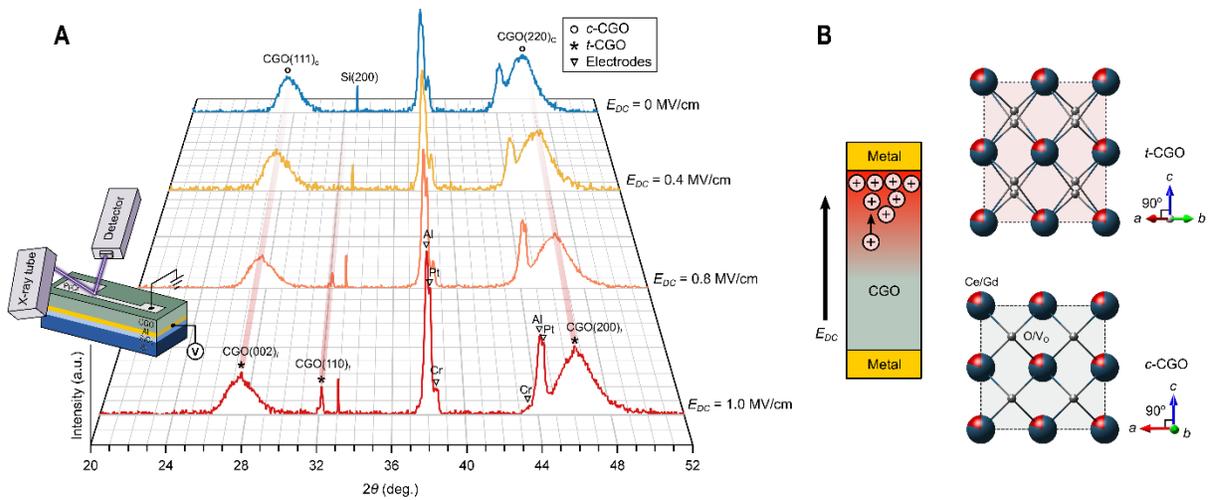

**Fig. 4. Impact of V$_O$ redistribution in centrosymmetric fluorite CGO film.** **(A)** XRD $2\theta$ patterns of the polycrystalline CGO film under various *in-situ* E$_{DC}$ applications (0, 0.4, 0.8, and 1.0 MV/cm). Schematic shows the *in-situ* XRD measurement setup using a lab-source X-ray ($\lambda$ = and top electrode area (~10 %) at the surface which was connected to the applied electric fields. Under $E_{DC} \geq 0.8$ MV/cm, new diffraction peaks were visible at $2\theta = 28.3°$, $32.6°$, and $46.9°$, which are assigned to (002), (110), and (200) reflections of a tetragonal phase of CGO. **(B)** Schematics for the phase transition of CGO from cubic to tetragonal phase through the field-induced redistribution of mobile positively charged V$_O$ (+).

**Materials and Methods**

**Film deposition:** CGO (thickness of ~1.25 – ~1.80 µm) and YSZ (thickness of ~0.3 µm) films were deposited at room temperature by using an RF magnetron sputtering (RF power: 200W, Ar gas flow: 15 sccm, and working pressure: 1 - 2 × $10^{-2}$ mbar) and a pulsed laser deposition (a 248 nm KrF Excimer laser, laser fluence: ~2 J/cm$^2$, repetition rate: 3 Hz, and working pressure: ~1 × $10^{-6}$ mbar). CGO (Gd: 20 %) and YSZ (Y: 8 %) ceramic targets were used for film deposition. Bottom Al electrodes (100 - 150 nm) on $SiO_2$ (200 nm)/Si(100) substrates were deposited using a DC sputter prior to film deposition, and top electrodes, Pt/Cr (Pt: ~100 nm and Cr: 10 nm) and Al (~150 nm), were deposited after film deposition. Microstructural properties, crystal structure, and elemental composition of the deposited polycrystalline films were determined by X-ray diffractometer (Bruker D8 advanced X-ray diffractometer, X-ray wavelength: λ = 1.54056 Å), atomic force microscopy (Cypher VRS), transmission electron microscopy (Double Cs-corrected ThermoFisher Scientific Titan Themis 60-300), and energy dispersive X-ray spectroscopy (EDX) performed at 200kV in the ThermoFisher Scientific Osiris microscope equipped with a Super-X detector at the EMAT laboratory. In the chemical analysis of the sputtered CGO films, Cr impurities of < 5 % were observed (error range of 1 - 2 %). To clarify the influence of such an impurity effect on the electromechanical properties of CGO, we deposited ~350 nm-thick CGO films on the same Al (100 nm)/$SiO_2$/Si substrates at RT using a pulsed laser deposition (PLD) technique. We confirm the large induced piezoelectric effect in the PLD-deposited CGO films without Cr/other impurities. Note that $Cr^{+3}$ and other acceptor dopants would, like Gd, create Vo in the film. (Fig. S15).

**Electromechanical characterizations:** Electric field-mechanical displacement responses of the films were recorded by implementing an advanced linear variable differential transformer (LVDT) and fiber-optic measurements. The LVDT sensor consists of a stationary element (including a primary magnetic coil, secondary coils symmetrically spaced with respect to the primary coil, and a movable core). Flat and hemispherical sample stages were electrically grounded. The LVDT sensor was equipped with two commercial lock-in amplifiers (SR830 DSP), a voltage source meter (Stanford Research System, Model No. DS360), and a voltage amplifier (Trek, Model No. 609c-6). The voltage source meter was used to apply AC voltages or combined AC & DC voltages to the samples. The two lock-in amplifiers are separately connected to drive LVDT and to measure output signal amplitudes (and an oscilloscope is used to display output signals). The output signals were converted to mechanical displacement via the calibrated instrument sensitivity (1.55 × 10⁻⁶ m/V). In the photonic sensing system (MTI-2100 Fotonic Sensor), samples were mounted on an alignment stage and displacements were measured from the reflective sample surface (grounded top electrode) in noncontact mode. Electric voltage was applied to the bottom electrode, connected with a lock-in amplifier and a voltage source meter. Photonic sensor has a linear voltage response to a change of sample -sensor distance. The effective probe area is approximately 0.79 mm² and an instrument sensitivity was 7.565 × 10⁻⁶ m/V. During displacement measurements in both of the above systems, all of the real-time output amplitude signals were concurrently recorded by using an oscilloscope (Tektronix, MDO3014). Prior to sample measurements, external instrument offset voltages, electrical noise, and instrumental sensitivity were corrected and confirmed by measuring reference samples, *e.g.* standard quartz crystal and PZT samples, in both of the systems.

**Supplementary Text**

**Section 1: Determination of harmonic electromechanical susceptibility.**

In this work, the 1st harmonic electromechanical strain ($x_{33}$) and susceptibility ($d_{33}$) of the samples in out-of-plane capacitor geometry were defined as:

$$x_{33} = \Delta L/L = d_{33}E_3, \qquad (1.1)$$

where, $x_{33}$ is the electric field ($E_3$)-induced out-of-plane strain, $\Delta L$ is the electromechanical displacement in length, $L$ is the total thickness of samples, and $E$ is the applied electric field.

$$d_{33} = \Delta L/V_{in} = [V_{out}\xi]/V_{in}, \qquad (1.2)$$

where $V_{out}$ is the measured displacement amplitude in volt, $\xi$ is the instrument sensitivity (1.55 × 10⁻⁶ m/V and 7.565 × 10⁻⁶ m/V for LVDT and photonic sensing system, respectively), and $V_{in}$ is the applied electric voltage amplitude. The electromechanical susceptibility (the piezoelectric coefficient) of the first harmonic is expressed by $V_{out}$ and phase lag, $\varphi$, in a function of applied frequency, $f$:

$$d_{33} = \frac{\xi V_{out} e^{i(\omega t - \varphi)}}{V_{in} e^{i\omega t}} = \xi \frac{V_{out}}{V_{in}} e^{-i\varphi} = \xi \frac{V_{out}}{V_{in}}(\cos\varphi - i \cdot \sin\varphi), \qquad (1.3)$$

where $\omega = 2\pi f$ is the angular frequency, $t$ is time, and $\varphi$ is the phase delay of the output response with respect to the applied AC electric field.

For the electrostrictive response of the samples, the 2nd harmonic electromechanical susceptibility, $M_{33}$ (m²/V²), of the samples in the out-of-plane capacitor geometry were determined as

$$x_{33} = \Delta L/L = M_{33}E_3^2, \qquad (1.4)$$

$$M_{33} = \frac{LV_{out}\xi}{V_{in}^2}. \tag{1.5}$$

The induced 2nd-order electromechanical displacements in time is expressed by a function:

$$\Delta L = \frac{M_{33}V_{in}^2\xi}{L}\sin^2(\omega t - \varphi). \tag{1.6}$$

$$\Delta L = \frac{M_{33}V_{in}^2\xi}{L}\left[\frac{1}{2} - \frac{1}{2}\cos(2\omega t - 2\varphi)\right]. \tag{1.7}$$

Thus, a pure 2nd-harmonic displacement in time is:

$$\Delta L = -\frac{M_{33}V_{in}^2\xi}{2L}\cos(2\omega t - 2\varphi). \tag{1.8}$$

For clamped thin films, the thickness strain $x_3$ is given by the modified $M$ coefficients:

$$x_{33} = [M_{33} - \frac{2s_{31}}{(s_{11} + s_{21})}M_{31}]E_3^2, \tag{1.9}$$

where $s_{ij}$ are components of the elastic compliance tensor of the film and $M_{31}$ transverse electrostrictive coefficient. For simplicity and without a loss of generality, eq. (1.4) is used in this work, with $M_{33}$ representing both contributions between the brackets of eq. (1.9), i.e. the effective $M_{33}$.

**Section 2: Capacitance measurements for the CGO samples**

The charge density, $D$ (C/cm²) was determined from the capacitance $C$ of the sample, as measured in a series RC circuit. The impedance Z of the circuit was excited by an AC voltage $Ve^{i\omega t}$ and the following relations are used for calculation:

$$V = IZ, \tag{2.1}$$
$$Z = R_{ex} + 1/(i\omega C), \tag{2.2}$$

where $I$ is the current, and $R_{ex}$ is the external resistance (10 - 100 Ω) of the series circuit. By substituting eq. (2.1) to eq. (2.2), the output voltage ($V_{out}$) of the circuit is obtained when a voltage ($V_{in}$) is applied as:

$$V_{out} = R_{ex}I = V_{in}R_{ex}/[R_{ex} + 1/(i\omega C)]. \tag{2.3}$$

Introducing a condition of $R_{ex} \ll 1/(\omega C)$ with $R_{sample}/R_{ex} \ll 0.01$ (see Fig. S3b), the capacitance of the RC circuit can be expressed by

$$C = V_{out}/(V_{in}R_{ex}i\omega). \tag{2.4}$$

The capacitance of the samples is described by

$$C = \frac{\varepsilon_0\varepsilon_r A}{L}, \tag{2.5}$$

where $\varepsilon_0$, $\varepsilon_r$, and A are the vacuum permittivity and the relative permittivity of the materials, and the measured electrode area, respectively. The total charge density of sample is obtained by combining eqs. (2.4) and (2.5):

$$D = \varepsilon_0 \varepsilon_r E = CV_{in}/A. \qquad (2.6)$$

**Section 3: Electric field-enhanced ionic defect dynamics and dielectric permittivity**

When the electrostriction is generated in a nonpolar material by electric field ($E$), the electric field-induced electrostrictive strain, $x$, can be expressed as $x = ME^2$. The polarization ($P$) of the material is induced by the applied $E$ and can be given as $P = \varepsilon E$. Combining the field-induced electrostriction and polarization, the polarization-induced electrostrictive strain can be defined as $x = QP^2$ (*17*).

In the relation, $M_{ijmn} = Q_{ijkl}\varepsilon_{km}\varepsilon_{ln}$, it was found that the dielectric constants $\varepsilon_{nm}$ of electrostrictive materials, such as, *e.g.*, (0.9)PMN-(0.1)PT, are nonlinear and staturate with applied electric fields. This indicates that $M$ is not constant when $\varepsilon$ is field-dependent and dispersive. Such a dependence is indeed observed in this work. Theroetical considerations as well as experiments show that $Q \propto 1/\varepsilon$ and thus $M \propto \varepsilon$. The well-known strong (inorganic) electrostrictive materials do not excel because of a large electrostrictive coefficient $Q$, but because of a large polarizability. The polarization ($P$), induced by an electric field ($E$) is large, which implies that the derivative $\varepsilon_{ij} = \frac{\partial P_i}{\partial E_j}$, *i.e.*, the dielectric response may also be large. In order to address the origin of large electro-mechanical properties of CGO (*e.g.* giant electrostriction and piezoelectricity), it is therefore primordial to indentify the origin of large permittivity and large induced polarization.

CGO has a cubic flourite structure (space group: *Fm-3m*) at RT and it cannot intrinsically exhibit a spontaneous polarization. The doped ceria, however, has many unit cells with broken symmetry because of the dopants (20 % Gd on cerium sites), the compensating oxygen vacancies (5 % of oxygen sites), and additional oxygen vacancies (*y*) which stem from a reducing process step in the material synthesis (*14*). It is believed that the liberated electrons are not entirely free, but trapped into the 4*f* states of Ce$^{3+}$. The complete formula thus must be written in the Kröger-Vink notation as $Ce^{\times}_{0.8-2y}Ce'_{2y}(Gd'_{Ce})_{0.2}O^{\times}_{1.9-y}([V_O^{\bullet\bullet}])_{0.1+y}$. The oxygen vacancies are mobile and the activation energy of their hoppings is about 0.45 – 0.70 eV in undoped CeO$_{2-x}$., stronlgy dependent on atmosphre conditon and microstructure (*19, 27-30*). This value increases to 0.75 eV in ceria doped with 20 % Gd as the $V_O^{\bullet\bullet}$ are "trapped" close to the Gd point defects (*14*). In addition, the dopants obstruct the easy path in the "*oxygen channels*" due to their larger size. By charge neutrality, there are twice as many Gd point defects as V$_O^{+2}$. The vacancies obtained by reduction can be trapped by Ce$^{+3}$ up to another 5 % of oxygen sites.

By introducing 20 % Gd into Ce sites, almost every unit cell (u.c.) has an oxygen defect ($V_O$ density of $2.52 \times 10^{21}$ cm$^{-3}$). It was found that, in thermal equilibrium (low temperatures below 600 °C), the preferred site of an oxygen vacancy is the next-next neighbour (*n.n.n*) site of Gd (*19, 27*). If Gd is placed in the coordinate (0,0,0) as in standard unit cell definition, the next oxygen sites are <1/4, 1/4, 1/4>. The n.n.n site are obtained by a basic translation of the fcc lattice, such as (1/2, 1/2, 0) yielding <1/4, 1/4, 3/4>. This next-next neighbor distance, $d$, is 4.48 Å [= 5.4179 Å ($a_{bulk}$) × 0.829] in CGO (Gd 20 %). Assuming that most vacancies are in such positions, we can identify a high density of dipoles, $qd$, where the charge, $q$, is one unit charge, $e$. Defining the average concentration, $c$, of Gd sites per unit cell to be 0.8 (4 cations per u.c. × 0.2), we obtain a potential total polarization, $P_s$, as:

$$P_s = \frac{c}{2} a^{-3} qd = 0.18 \text{ C/m}^2 \tag{3.1}$$

The orientation of this polarization is not stable and thus must be forced by an external field ($E_{DC}$). The reorientation of the polarization is effectuated by oxygen ion hopping. The basic polarization reorientation mechanism is based on one or several hops of oxygen vacancies, which require an activation energy of $\phi$=0.32 to 0.75 eV depending on the distance to the next Gd as shown in Fig. S10 (*19*). Typically, we treat this hopping as an ordinary diffusion process in a weak electric field. In the experiments carried out here, the applied large clamping fields are not weak, one has to evaluate the jumping probabilities. We emphasize the possibility explicitely for both directions and/or against the electric field:

$$W(+) = e^{-\phi/kT} e^{+qEh/2kT}, \tag{3.2}$$
$$W(-) = e^{-\phi/kT} e^{-qEh/2kT}, \tag{3.3}$$
$$W(\text{forward}) = e^{-\phi/kT} \left( e^{+qEh/2kT} - e^{-qEh/2kT} \right). \tag{3.4}$$

These probabilities describe the success rate for the jumps attempted $v$ times per second. The parameter $h$ is the hopping distance, typically by half a lattice constant (~2.7 Å). For the attempt rate, we use a typical phonon frequency of ~$10^{11}$ Hz, which was experimentally determined by means of diffusion studies in other works. In absence of an electrical field, the effective hopping rate at 300 K is:

$$R_0 = v e^{-\phi/kT}, \tag{3.5}$$
$= 2.5 \times 10^{-2}$ Hz ($\phi = 0.75$ eV) or 400 Hz ($\phi = 0.5$ eV) or 13 kHz ($\phi = 0.41$ eV)

This implies the corresponding relaxation times, $\tau_0 = 1/R_0$, of 40 s, 2.5 ms, and 76.8 µs, respectively. The effective hopping rate in the direction of an electric field can be written as:

$$R(E) = v e^{-\phi/kT} \left( e^{+qEh/2kT} - e^{-qEh/2kT} \right). \tag{3.6}$$

The single polarization average is written by:

$$<p(t)> = qd(1 - e^{-R(E)t}), \tag{3.7}$$

$$P(t) = \frac{\frac{c}{2}\left(1-\frac{c}{2}\right)}{a^3} <p(t)> \quad (3.8)$$

Using the time-dependent polarization, the dielectric constant can be derived as:

$$\varepsilon_r = \frac{\partial P}{\partial E} = \frac{\frac{c}{2}\left(1-\frac{c}{2}\right)}{a^3} qde^{-R(E)t}vte^{-\frac{\phi}{kT}}\frac{qh}{2kT}\left(e^{+\frac{qEh}{2kT}} + e^{-\frac{qEh}{2kT}}\right) \quad (3.9)$$

$$= \frac{\frac{c}{2}\left(1-\frac{c}{2}\right)}{a^3} vqde^{-R(E)t}\frac{t}{\tau_0}\frac{qh}{2kT}\left(e^{+\frac{qEh}{2kT}} + e^{-\frac{qEh}{2kT}}\right).$$

By integrating time, $t$, over a sinusidal period, $1/f$ or $2\pi/\omega$, eq. (4.8) can be given as:

$$\varepsilon_r = \frac{\frac{c}{2}\left(1-\frac{c}{2}\right)}{a^3} vqd\frac{qh}{2kT}\frac{2\pi}{\omega\tau_0}\left(e^{+\frac{qEh}{2kT}} + e^{-\frac{qEh}{2kT}}\right)e^{-R(E)t} \quad (3.10)$$

By introducting parameters, $t = 0$ (far from saturation), $T = 300$ K, $\phi = 0.41\pm2$ eV (experimentally obtained, see Fig. S10), and $c = 0.8$, giant dielectric permittivity in CGO (20% Gd) can be induced with lowering frequency as shown in Fig. S10. Our resuls clearly show the dielectric permittivity of CGO strongly relies on the activation energy of ionic defect hopping and its rate (frequency dependence) under the applied $E_{DC}$. The proposed ionic diffusion model and the corresponding variations in the dielectric permittivity of CGO give excellent agreement with the experimental results. Therefore, we affirm that the static field-enhanced ionic hopping/conduction drives large frequency-dependent dielectric permittivity, which helps generating the correlated large piezoelectricity. Note that this hopping model does not include phase transition and electric field-induced $V_O$ separation on large-scale distance of the CGO films. Therefore, this model is valid for modelling high-frequency dielectric constant, *i.e.*, the value of several hundred measured at 1 kHz.

**Section 4. X-ray diffraction measurements with *in-situ* electric field applications.**

Structural variations of the polycrystalline CGO (Gd 20%) films were investigated using *in-situ* XRD (Bruker D8 advanced X-ray diffractometer, λ (Cu-*Kα* radiation) = 1.54056 Å) under different dc electric fields (0 - 1 MV/cm). The samples were measured, which were the exactly same samples used for our electromechanical tests. The electrode (Pt/Cr) area of the films is about 15 % to the total surface area of the samples. The electric fields were applied from top electrode (Pt/Cr) to bottom electrode (Al) while measuring XRD. To acquire sufficient XRD intensity for the field-induced structural variations of the films, long scan time for each angle step (5 - 10 sec/step) were employed for XRD measurements with fine angle increments (0.005° – 0.01°). All of the collected diffraction spectra were further calibrated by referring to the multiple diffraction of the forbidden Si (200) at $2\theta = 33°$ (*31*). The pristine CGO films deposited at room temperature are polycrystalline and predominantly show broad diffraction CGO (111) and (220) peaks at $2\theta = 27.49°$ and $46.12°$, respectively. These peak angles of the pristine films are lower than those of ideal CGO (Gd 20%) ($2\theta = 28.47°$ and $47.36°$). This indicates that the film has a larger lattice constant (5.61 Å) compared to that of bulk CGO (Gd 20%) (5.425 Å). This is due to additional amount of oxygen vacancies in the films around 8% (*14*), which can result in the Vo-induced chemical expansion of the pristine film due to the valence change from $Ce^{4+}$ (97 pm) to $Ce^{3+}$ (114 pm) (*20*).

Remarkably, XRD clearly shows the appearance of a new peak at $2\theta = 32.16$ when the electric field applied to the sample increases above $E_{DC} = +0.8$ MV/cm. It is likely that additional peaks appear at $27.78°$ and $46.03°$, but are hidden by overlapping cubic reflections. The peaks may be present at lower fields but are below detection limit. This directly indicates appearance of a field-induced phase transformation in the CGO film. Phase transitions (among three solid polymorphs, cubic (space group *Fm-3m*), tetragonal (*P4₂/nmc*), and monoclinic (*P2₁/c*) symmetries) have been often observed in fluorites ($ZrO_2$, aliovalent cation-stabilized $ZrO_2$, and $CeO_2$) by controlling doping, temperature and applying large electric field at relatively high temperature (*22, 25, 32, 33*). For example, *Zhu et al.* reported that the effect of $V_O$ in $CeO_2$ which stabilizes a tetragonal phase (*P4₂/nmc*) via charge transfer ($V_O \rightarrow Ce^{3+}$) at temperatures below 100 °C (*25*). Hence, $V_O$-induced phase transition in the CGO films can occur and be promoted when the density of $V_O$ at the cathode increases by high applied DC&AC electric fields. With the tetragonal phase within the CGO film (Fig. S16), we obtained the tetragonal lattice parameters are $a_{(T)} = b_{(T)} = 3.94$ Å and $c_{(T)} = 6.41$ Å.

In order to better apprehend this transformation, one has to be aware of the usual set of elementary vectors describing the Bravais FCC lattice. The FCC lattice is a non-primitive cubic lattice. It contains the ones of the primitive unit cell $[\frac{a}{2},\frac{a}{2},0]$, $[\frac{a}{2},0,\frac{a}{2}]$, $[0,\frac{a}{2},\frac{a}{2}]$, spanning the primitive unit cell with one formula unit, and the ones of the cubic frame ([a,0,0], [0,a,0], [0,0,a]), which includes 4 formula units ($Z = 4$, where $Z$ is the number of formula units in the crystallographic unit cell). A tetragonal unit cell can be equally well defined with the base plane spanned by the primitive unit vectors $[\frac{a}{2},\frac{a}{2},0]$, $[\frac{a}{2},-\frac{a}{2},0]$ and the basic translation [0,0,*a*] along the 4-fold axis perpendicular to the base plane. We still need to add a third primitive unit cell vector $[\frac{a}{2},0,\frac{a}{2}]$. This non-primitive, tetragonal unit cell has $Z = 2$, and half the volume of the standard FCC unit cell ($a^3$) (*24*). The unit cell dimensions of the observed new tetragonal phase have to be compared with the tetragonal choice of the FCC unit cell axes, *i.e.*, $\sqrt{2}a(t) = \sqrt{2}b(t) = 5.57$ Å, $c(t) = 6.42$ Å. Therefore, the corresponding strains in crystallographic *a* and *c* axes are $(\sqrt{2}a_{(T)} - a_{(C)})/a_{(C)} \approx -0.7\%$ and $(c_{(T)} - c_{(C)})/c_{(C)} \approx +14.5\%$, respectively.

## Section 5. Frequency dependence of *d/M*

Three points need to be recalled for ensuing discussion. First, materials coefficients are derivatives of strain, and polarization ($\varepsilon = \partial P/\partial E, d = \partial x/\partial E, Q = \frac{1}{2}\partial^2 x/\partial P^2$); therefore, it is the change of polarization and strain with field that matters for a large response and not their absolute values. Second, the field induced transformation into tetragonal phase is partial, *i.e.*, during application of the field of sufficient magnitude and after sufficient time, the sample exhibits mixed cubic and tetragonal phases. Third, the field induced phase transformation appears to be reversible.

During electrostrictive measurements (Fig. 1 in the main text), only moderate $E_{AC}$ is applied. This field can at best cause partial phase transformation and associated strain near the field peak, when $E_{AC}$ reaches critical field needed for phase transformation, as the field is cycled between $-E_{AC}$ and $+E_{AC}$. At frequencies above ~10 Hz, the phase transformation is limited under assumption that it is assisted or enabled by $V_O$ migration because defects cannot follow the field easily. The electrostriction coefficient *M* is then essentially controlled by $\varepsilon$ and *Q* of the cubic phase, both of which are mostly due to the intrinsic contribution and some polarization due to the limited field-induced charge migration and hopping. As the frequency decreases, motion of $V_O$ becomes more pronounced, contributing both to $\varepsilon$ and to the strain from the phase transformation. This process defines the rate-dependent electrostrictive coefficient.

On the other hand, piezoelectric measurements are made under strong $E_{DC}$ and a small $E_{AC}$. Under $E_{DC}$ of sufficient strength, the sample consists at all times of a mixture of tetragonal and cubic regions. Under $E_{DC}$ the energy barrier among these regions must be low, so that even a relatively weak $E_{AC}$ fields can reversibly switch between the cubic and tetragonal phases in some volume fraction of the sample, if enough time is given for displacement of $V_O$ that increases susceptibility for the transformation. This was directly observed by a concurrent measure of $|d_{33}|$ and $|M_{33}|$ under a high $E_{DC}$ as shown in Fig. 3D. At frequencies above about ~10 Hz, the weak $E_{AC}$ cannot move $V_O$ fast enough to promote further oscillating phase transformations. The field-induced *d* is thus governed by the values of *Q* and $\varepsilon$, which are representative of the mixture of the cubic and tetragonal phases. The polarization electrostrictive coefficient and permittivity are not expected to change much (less than an order of magnitude) between different phases (*34*). This means that in this high frequency range the *d* coefficient is essentially controlled by the same intrinsic relationships as *M*, leading to a good agreement between the measured *d* and the value estimated from $2ME_{DC}$. The small discrepancy between expected and measured values at high frequency (1 kHz) can be explained by the differences in values of the *Q* coefficients in the tetragonal and cubic phase and weak extrinsic contributions to $\varepsilon$, which will be different under conditions used for measuring *d* and *M*. At low frequencies, the rate of change of field is slow enough that motion of $V_O$ can help reversible cubic-to-tetragonal (and back) transformation during $E_{AC}$ cycling, leading to a huge contribution to the apparent piezoelectric strain.

Furthermore, the direct longitudinal piezoelectric effect ($D_3 = d_{33}\sigma_3$ or $E_3 = -g_{33}\sigma_3$, where $\sigma$ and *g* are applied pressure and direct piezoelectric voltage coefficient, respectively) (*35*) should be measurable by application of a dynamic pressure once the required fundamental condition, macroscopic symmetry breaking is achieved by externally applied direct electric field $E_{DC}$. Such an experiment requires in measuring either small *D* (*d* coefficient) or small *E* (*g* coefficient) in presence of a large $E_{DC}$ while simultaneously applying dynamic stress. The stress may also interfere with the conditions needed to achieve the phase transition.

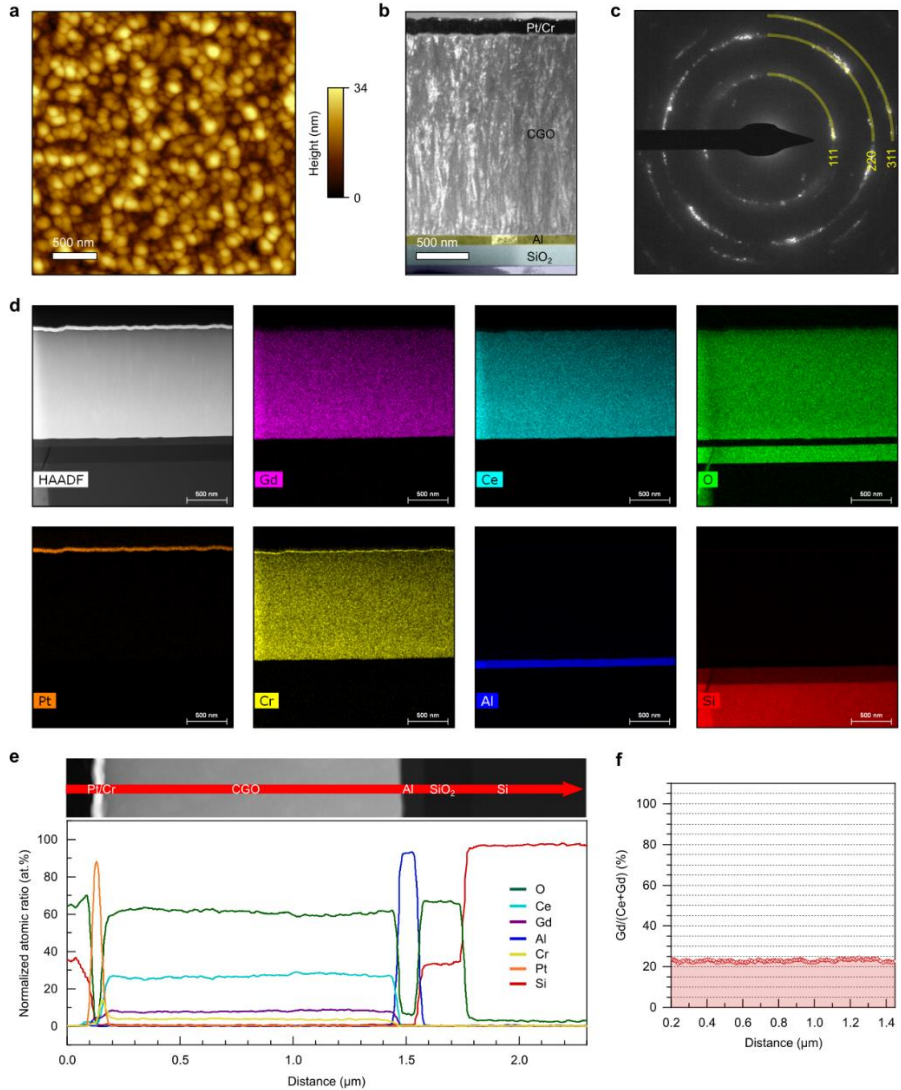

**Fig. S1. Surface morphology, microstructural properties, elemental mappings, and line profiles of CGO film samples.** (**a**) Atomic force microscopy topography image (3 × 3 μm$^2$) of the surface of a CGO film. (**b**) A cross-sectional dark field transmission electron microscopy image of the Pt/Cr/CGO (~1.8 μm-thick)/Al/SiO$_2$/Si sample. (**c**) Electron diffraction patterns of the polycrystalline CGO film. (**d**) EDX elemental mapping of a CGO film (~1.25 μm-thick) sample. For comparison, the EDX mapping images of different elements (Gd, Ce, O, Pt, Cr, Al, and Si) are presented separately. (**e**) EDX line profiles of the CGO film sample along the arrow, shown in the top cross-sectional high-angle dark field scanning transmission electron microscopy image. (**f**) The atomic ratio of Gd to Ce in the CGO film, determined to be 22.8±1.6 %. According to the EDX analysis, Cr impurities (~2 %) were found in the film layer, of which such contaminants are from deposition tools/external mechanical processing.

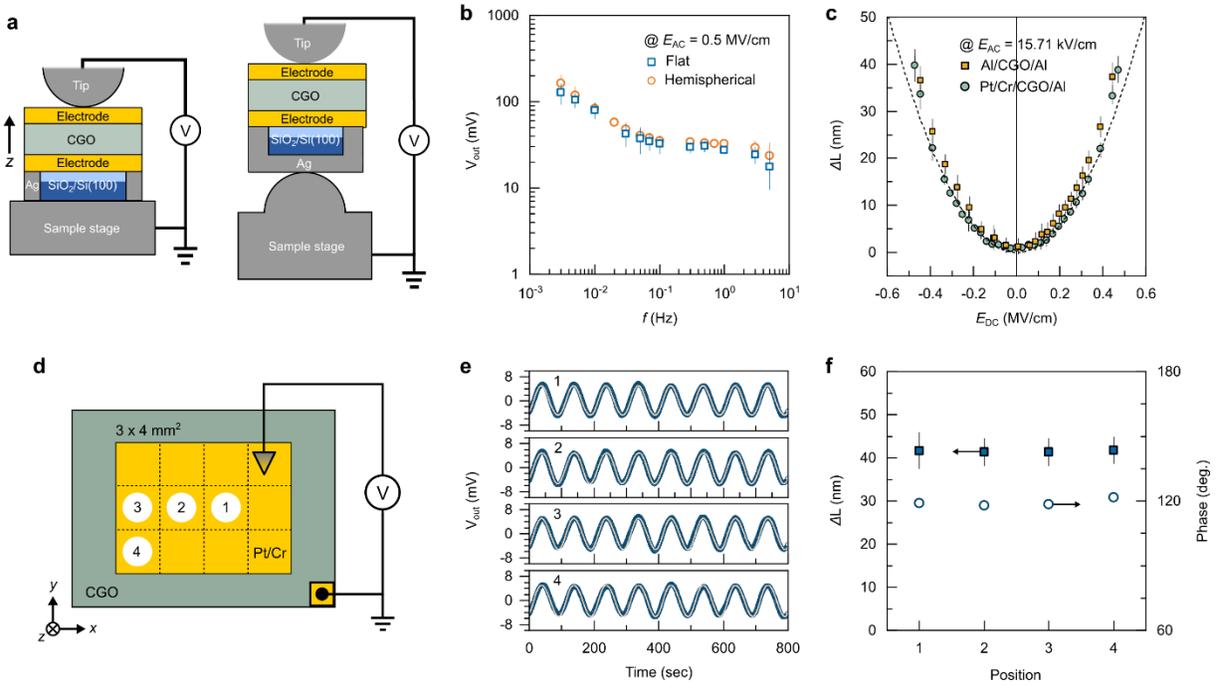

**Fig. S2. Verification of measurement artifacts.** (**a**) Schematics of different sample clamping methods for the sample measurements in LVDT. Two different sample stages, flat and hemispherical stages, were employed. A radius of the hemispherical sample stage is 2.5 mm, smaller than any of samples measured. (**b**) Out-of-plane displacement amplitudes ($V_{out}$) of the CGO samples as a function of frequency, determined by separately using the flat and hemispherical metal stages. (**c**) Variations in the 1st order electromechanical displacements of the CGO films with different top electrodes (Pt/Cr and Al), excited by simultaneously applying $E_{AC}$ = 15.71 kV/cm and $E_{DC}$ = +0.5 MV/cm. This indicates no significant asymmetric Schottky effect for the generation of large electromechanical responses of CGO films. (**d**) Schematic of different probe points on a CGO sample for mechanical measurements in the photonic sensing system. Samples were firmly mounted on a flat metallic stage by Ag paste. Top (Pt/Cr or Al) electrodes were connected to bottom (Al) electrode by a metal tip to apply electrical field. Electromechanical displacements for four different probe points (photonic sensor positions, denoted as 1, 2, 3, and 4) were recorded while fixing the tip contact under a constant electric bias apply ($E_{AC}$ = 15.71 kV/cm & $E_{DC}$ = -0.5 MV/cm). (**e,f**) The measured displacements and response phase of the probe points with respect to the applied $E_{AC}$. By comparing the above measurement setups and output responses in two instruments, we confirm that substrate bending is suppressed by gluing the sample to the flat stage, thus having no effect during the electromechanical measurements.

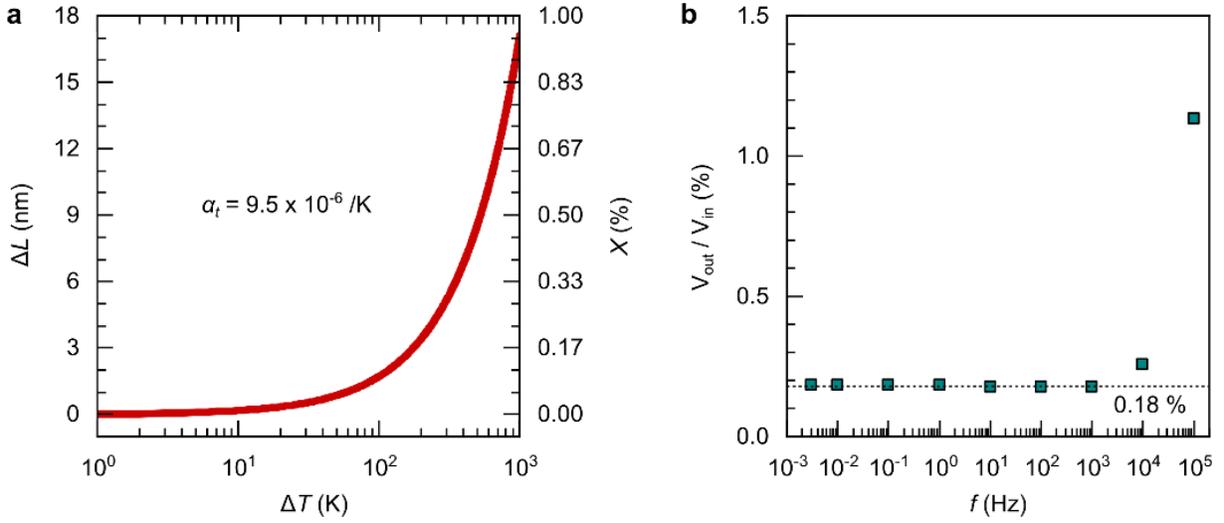

**Fig. S3. Expected thermally induced volumetric strain effect in CGO (Gd 20 %) and external resistor effect in series RC circuit.** (a) The thermal expansion of CGO films is predicted by a linear formula, $X = \Delta L/L = \alpha_t(\Delta T)$, where $\alpha_t$ and $T$ are the thermal expansion coefficient ($9.5 \times 10^{-6}$ K$^{-1}$) for the case of CGO (Gd 20 %) and temperature change, respectively (*36*). A temperature increase of < 40 °C was found during high-field applications ($E_{DC}$ ~0.8 MV/cm at 10 mHz) by directly attaching a thermocouple to the surface of the sample. (b) Output voltage ($V_{out}$)/input voltage ($V_{in}$) ratio in frequency during the electrical measurements. External resistors, $R_{ex}$, are added in series with the sample and voltage drop ($V_{out}$) is measured across $R_{ex}$ to determine electrical current and electro-mechanical responses of the sample. As the current across $R_{ex}$ can be a strong function of frequency, $V_{out}$ can also show a strong frequency dependence. To eliminate this effect, a constant very low $V_{out}/V_{in}$ ratio ($\omega R_{ex}C \ll 1$ %) was kept during the concurrent electrical and electromechanical measurements.

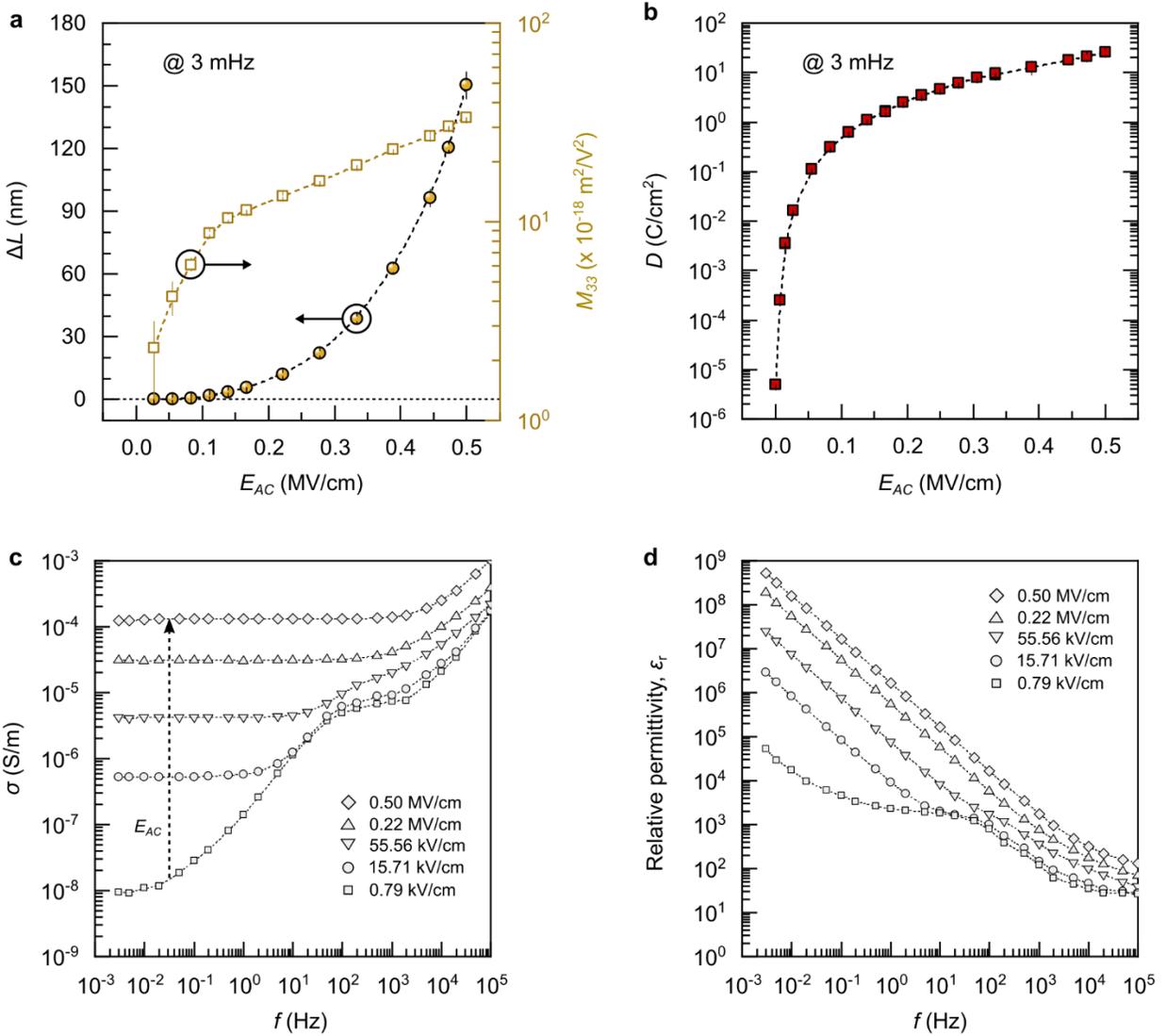

**Fig. S4. Variations in the electrostrictive response, charge density, conductivity, and dielectric permittivity of the CGO film as functions of applied $E_{AC}$ and frequency $f$.** (**a**) The 2nd order electromechanical displacements, $\Delta L$, and susceptibility, $M_{33}$, of the CGO film with $E_{AC}$ at $f = 3$ mHz. A large increase in the low-frequency $M_{33}$ of the film occurs when $E_{AC}$ increases. (**b**) The corresponding change in the charge density, $D$, of the film with $E_{AC}$. (**c**) The $f$-dependent electrical conductivity of the sample as a function of $E_{AC}$. (**d**) The relative dielectric permittivity of the CGO film as a function of $E_{AC}$ and $f$.

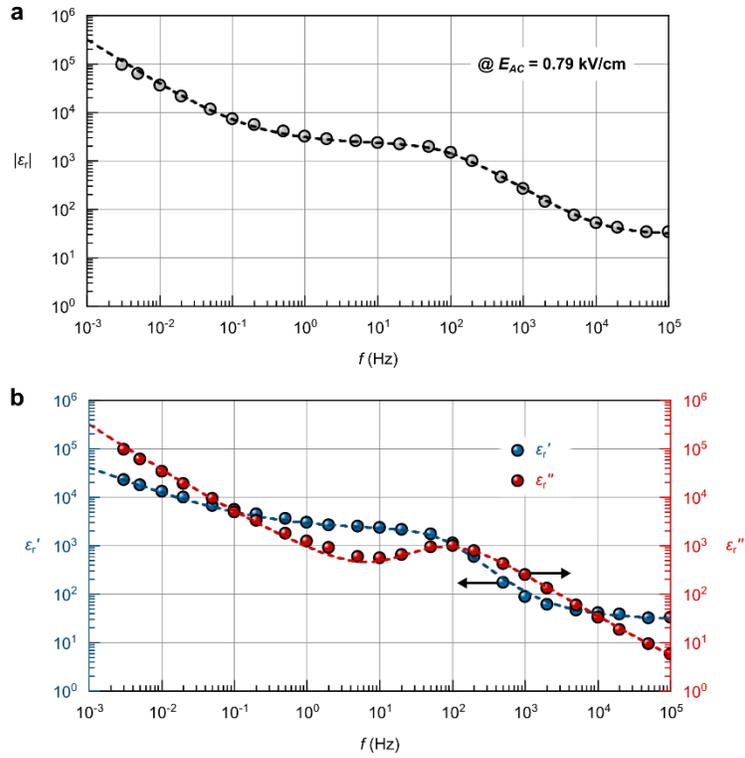

**Fig. S5. Complex apparent dielectric permittivity and relaxation characteristics of the CGO film.** (a) Modulus $|\varepsilon_r| = \sqrt{\varepsilon_r'^2 + \varepsilon_r''^2}$ of the apparent relative dielectric permittivity of the CGO film as a function of $f$, excited by $E_{AC} = 0.79$ kV/cm. The experimental dielectric permittivity (circles) were fitted by a modified complex dielectric relaxation model (dashed lines) incorporating multiple relaxation processes and the following conductivity terms (*37*). (b) The real ($\varepsilon'$) and imaginary ($\varepsilon''$) parts of the permittivity of the CGO film. The magnitudes of the $\varepsilon'$ and $\varepsilon''$ significantly increase with decreasing $f$.

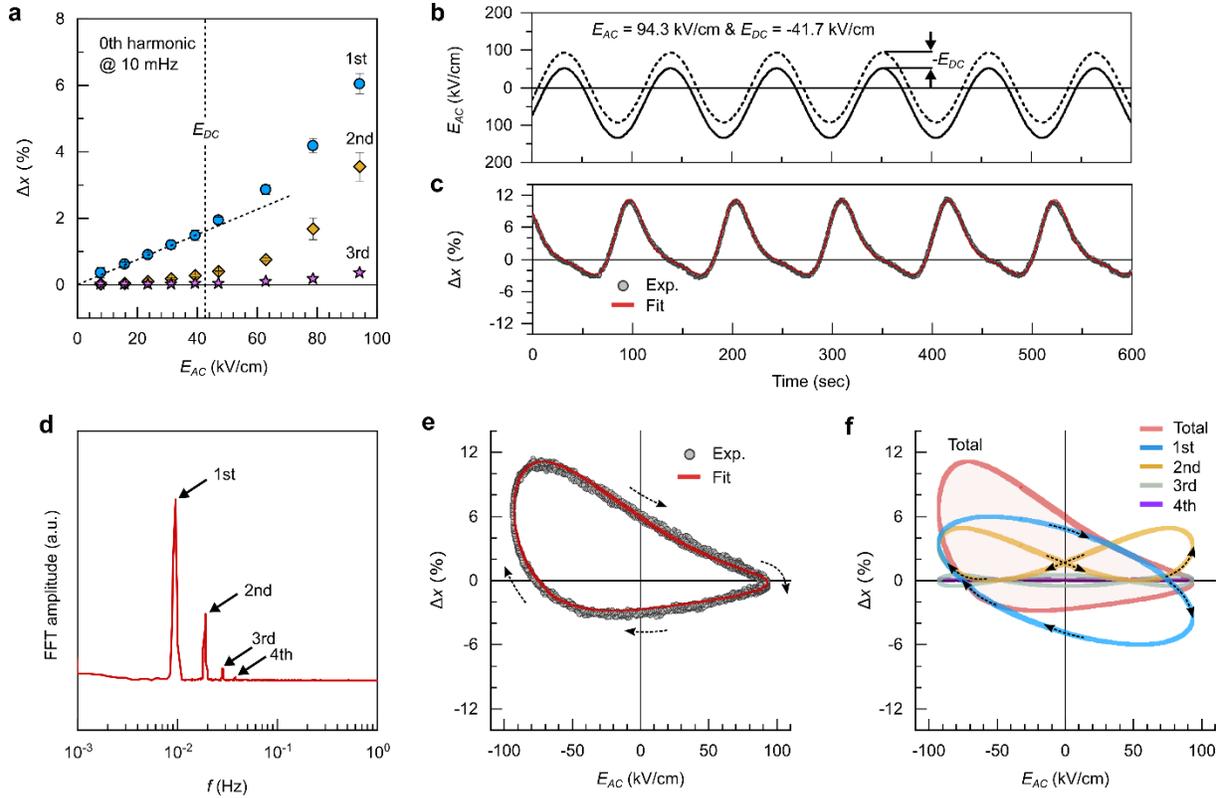

**Fig. S6. Nonlinear electromechanical responses of CGO film.** (**a**) The 1st, 2nd, and 3rd harmonic electromechanical strains of the film as a function of $E_{AC}$, measured at 10 mHz. (**b**) Applied combined $E$ ($E_{AC}$ = 94.3 kV/cm and $E_{DC}$ = -41.7 kV/cm) to the sample at $f$ = 10 mHz. (**c**) Spectral analyses of the measured strain (grey circle) of the film, fitted by multiple sinusoidal functions (red solid line), $x_{Tot.} = \Sigma\, x_{0,n}\sin(n\cdot\omega t - \varphi)$. (**d**) FFT analysis for the measured electromechanical signals in frequency. (**e**) Strain Vs $E_{AC}$ curve of the CGO film under $E_{DC}$ (-41.7 kV/cm). (**f**) A spectral analysis of strain distribution in the film during the field application, fitted by combining four (1st to 4th) harmonic components in total. Such high-order harmonics originate from the nonideal electrostriction in CGO (and most other materials) where the associated total ideal quadratic strain, $x = ME^2$ can be modified by expressing the coefficient $M$ as being field dependent, $M = m + nE + pE^2 + \cdots$, thus giving $x = mE^2 + nE^3 + pE^4 + \cdots$ (*17*). *Also, asymmetric/nonlinear strain Vs electrical field responses can be generated by instrumental voltage offsets and thus careful instrumental calibration is always required for sample measurements.*

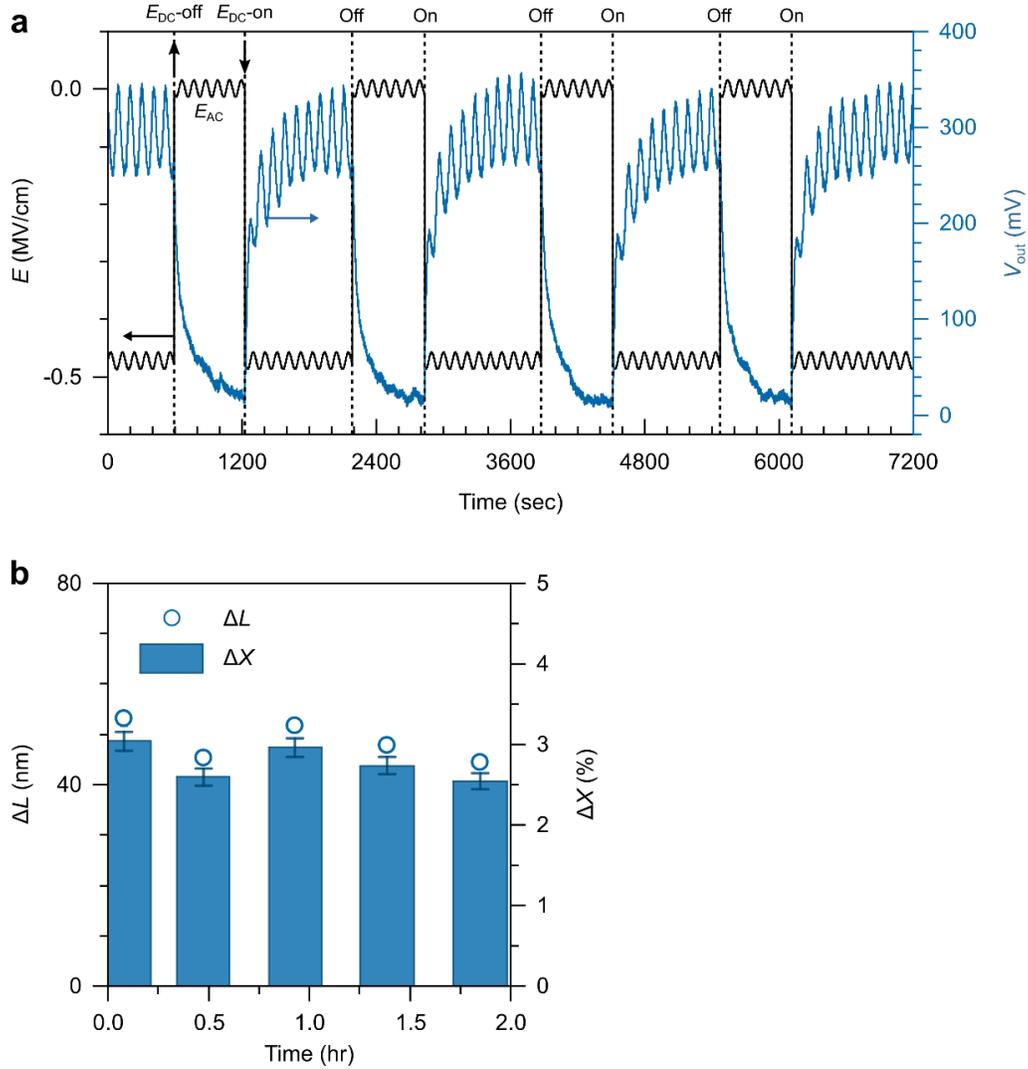

**Fig. S7. On-off control of the 1st harmonic electromechanical responses of the CGO film by simultaneously applied DC field.** (**a**) The 1st harmonic displacement amplitude ($V_{out}$) of the CGO film in time. The output amplitude is switched by on-and-off negative DC field ($E_{DC}$ = -0.5 MV/cm) while applying a constant AC field ($E_{AC}$ = 15.71 kV/cm). (**b**) The corresponding electromechanical displacements and strain of the film in time.

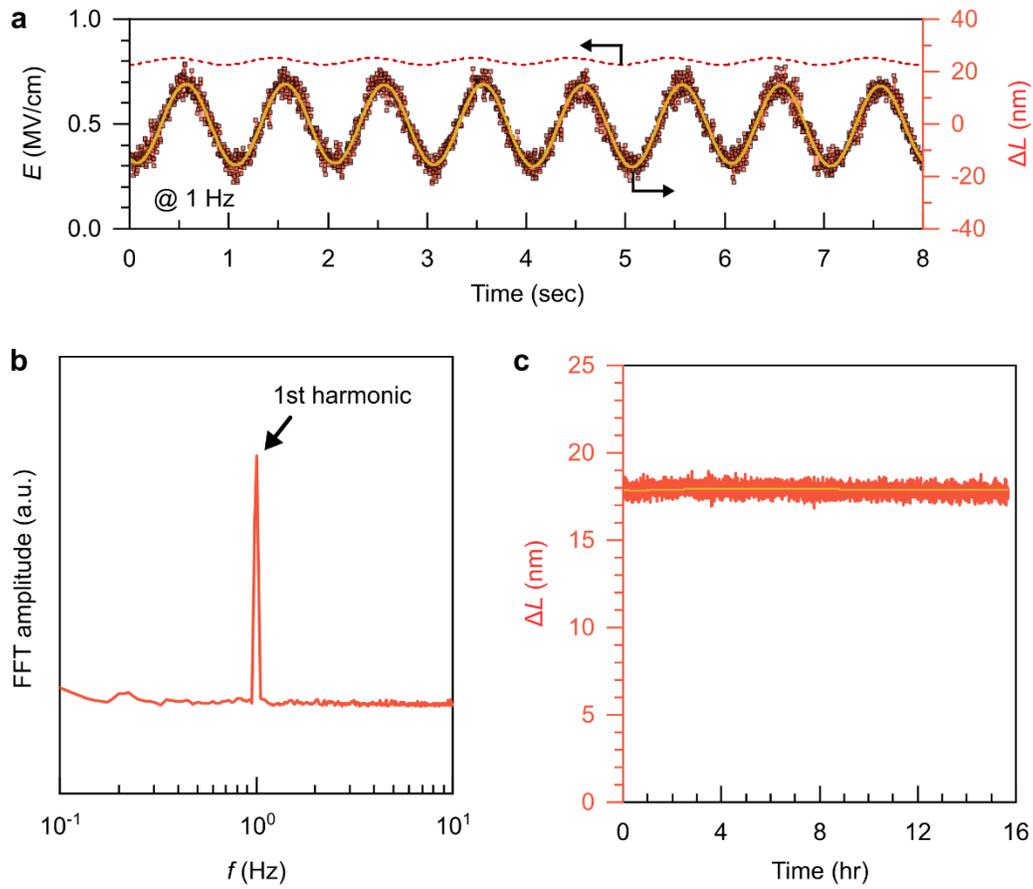

**Fig. S8. Endurance performance on the induced piezoelectric displacement of the 1.25 μm-thick CGO film.** (a) The piezoelectric displacement (max. 1st harmonic strain, $\Delta x$ ~1.8 %) of the CGO film in time, excited by a combined electric field [$E_{AC}$ (16 kV/cm) + $E_{DC}$ (0.8 MV/cm)] at 1 Hz. (b) The corresponding FFT amplitude spectrum of the 1st harmonic output electromechanical signals in $f$. (c) A long-time (> 15 hrs) output signal acquisition for the 1st harmonic piezoelectric displacement of the CGO film. This result shows excellent endurance of the piezoelectric performance of the CGO sample without significant signal degradation.

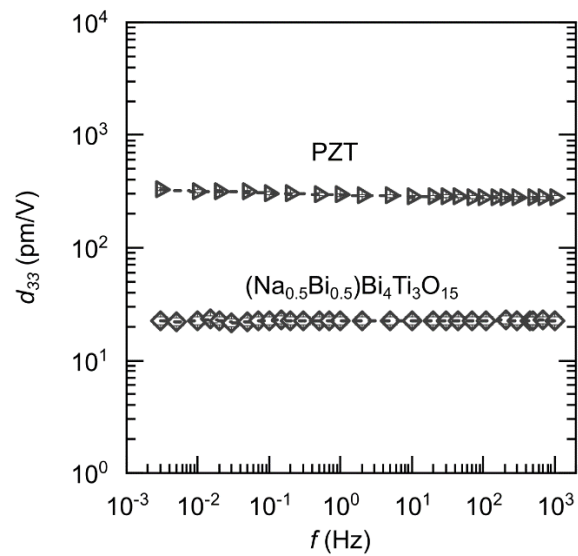

**Fig. S9. Frequency-independent piezoelectric susceptibility of PZT and $(Na_{1/2}Bi_{1/2})Bi_4Ti_3O_{15}$-based material.** Piezoelectric coefficients ($|d_{33}|$) of PZT and $(Na_{1/2}Bi_{1/2})Bi_4Ti_3O_{15}$ (PZ46-Ferropem) ceramics in the frequency range of 3 mHz – 1 kHz, measured by $E_{AC} = 1.25$ kV/cm.

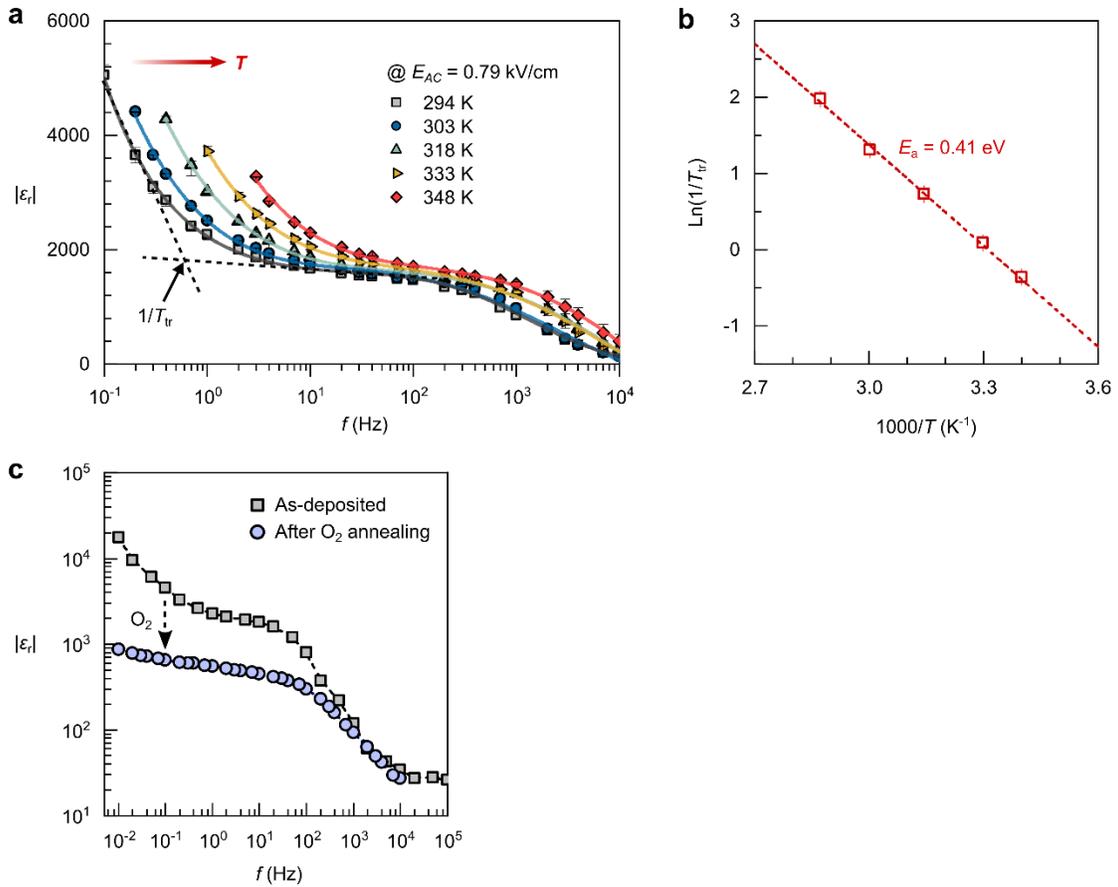

**Fig. S10. Dielectric dispersion of CGO film with temperature variations.** (**a**) Dielectric permittivity, $\varepsilon_r$, of the CGO film as a function of $f$, measured in the range of $T = 294 - 348$ K. Transient times, $T_{tr}$, for the characteristic dielectric relaxation in a CGO film were determined by linear extrapolation of the onsets to the plateaus. (**b**) an Arrhenius plot of $\mathrm{Ln}(1/T_{tr})$ in 1000/T. An activation energy of the $T_{tr}$ was determined to be $E_a = 0.41 \pm 0.02$ eV. (**c**) $\varepsilon_r$ of a CGO film as a function of frequency before and after $O_2$ annealing. The sample was annealed in $O_2$ flow at 400 °C for 30 mins. After annealing the permittivity decreases by more than an order of magnitude at low frequencies, indicating a large contribution of $V_O$ to the permittivity and therefore piezoelectric properties at low frequencies.

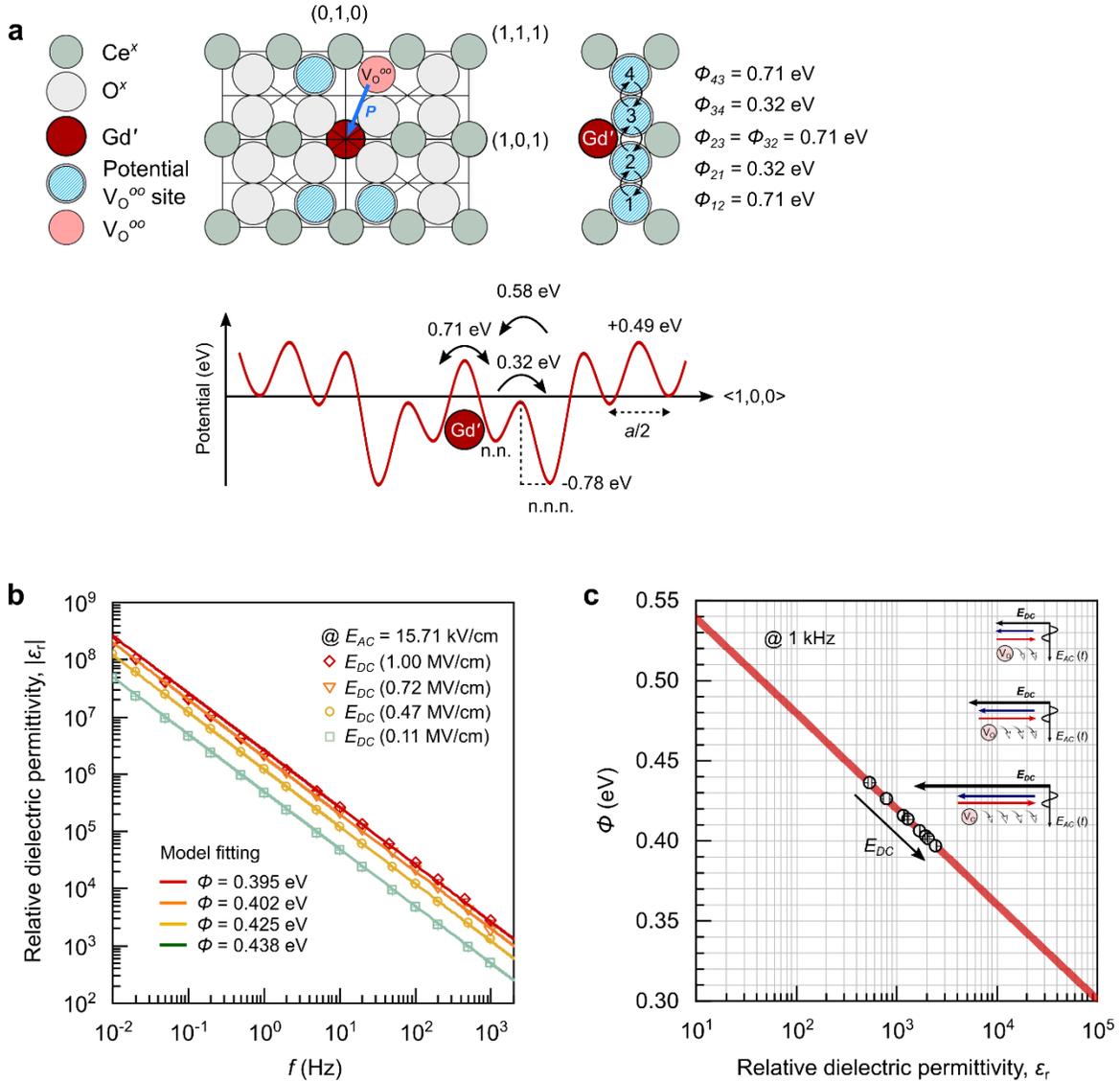

**Fig. S11. Calculations on defect hopping, polarization, and dielectric constant of CGO (Gd 20%) system.** (a) Schematics of charge polarization between Gd dopant at coordinate (0,0,0) and next neighboring $V_O$ at the coordinate (¼,¼,¼) in cubic fluorite CGO (Gd 20%) structure. In the right side, energetic description of $V_O$ jumping processes in a simplified 4-oxygen site model. In the lower side, different energy potentials of $V_O$ jumping at oxygen sites with respect to Gd site. This reveals preferential Vo hopping is along the next-next neighboring oxygen sites (¼, ¾, ¼). The values of oxygen hopping energies were referred to (*19*). (b) Fittings for the experimental frequency-dependent $|\varepsilon_r|$ of the CGO film, excited by simultaneous field application of a constant $E_{AC}$ and different $E_{DC}$ (0.11, 0.47, 0.72, and 1 MV/cm). The fittings are obtained using Eq. 3.10 with different values of $\Phi$. (c) Experimental high-frequency (1 kHz) dielectric permittivity and the corresponding calculated $\Phi$ of the CGO film with different $E_{DC}$.

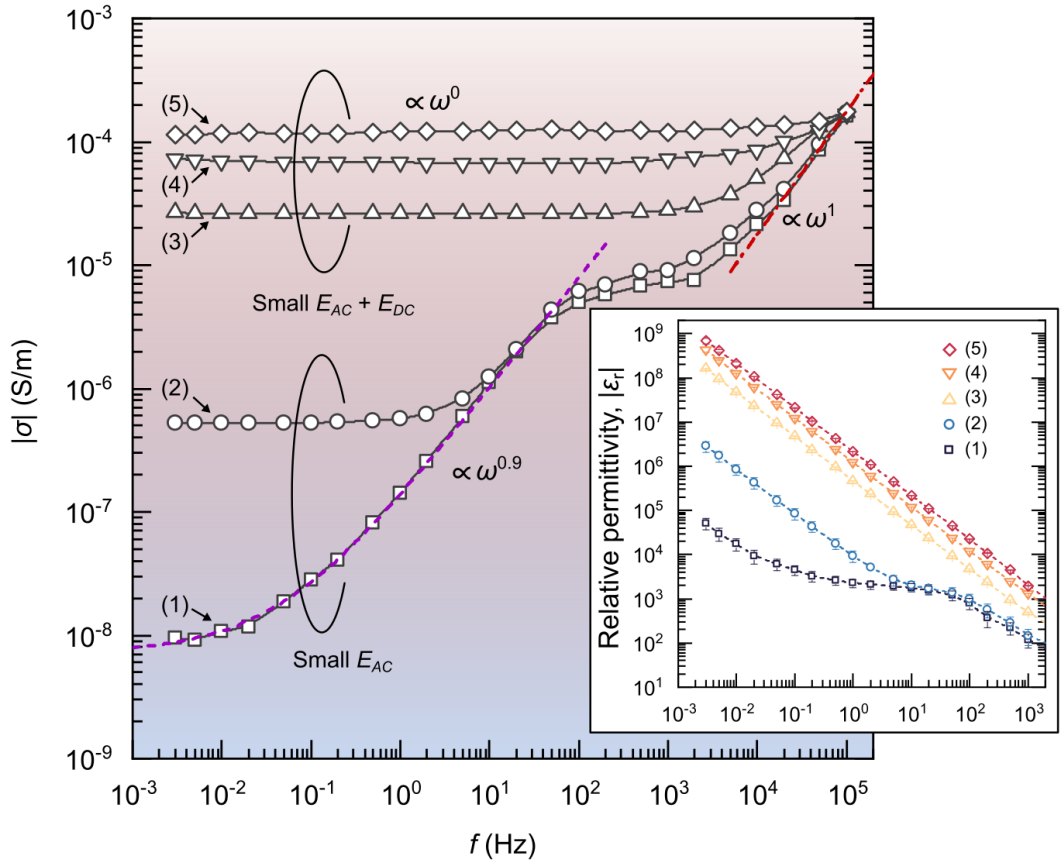

**Fig. S12.** Frequency-dependent electrical conductivity, $|\sigma| = \sqrt{\sigma'^2 + \sigma''^2} \approx \sigma'$, of a CGO (Gd 20%) film with different field excitations: (1) with $E_{AC} = 0.79$ kV/cm, (2) with $E_{AC} = 15.71$ kV/cm, (3) with $E_{AC} = 15.71$ kV/cm and $E_{DC} = 0.11$ MV/cm, (4) with $E_{AC} = 15.71$ kV/cm and $E_{DC} = 0.47$ MV/cm, and (5) with $E_{AC} = 15.71$ kV/cm and $E_{DC} = 0.72$ kV/cm. The dash line represents an ionic relaxation behavior, expressed by $\sigma(\omega) = \sigma_0 + A\omega^n$, where $\sigma_0$ is the *dc*-conductivity (low-frequency), $A$ is temperature-related term and the exponent, $n$, lies in the range $0 < n < 1$. The dashed line with $n = 0.9$ in relatively low frequency range (< below 1 kHz) represents a dispersive ionic diffusion (*38, 39*), while the line with $n = 1$ at high frequency range (> 1 kHz) represents a universal limiting conductivity (*40*). With simultaneous application of AC and DC fields to the sample, the *Jonscher*'s power law is absent ($n \approx 0$) and frequency-independent dc-conductivity occurs, indicating greatly enhanced long-range ionic diffusion/hopping process (*41*). The inset shows the corresponding logarithmic $f$ versus logarithmic $|\varepsilon_r|(\sqrt{\varepsilon_r'^2 + \varepsilon_r''^2} \approx \varepsilon_r'')$ for the excited CGO sample with different field applications (1 - 5).

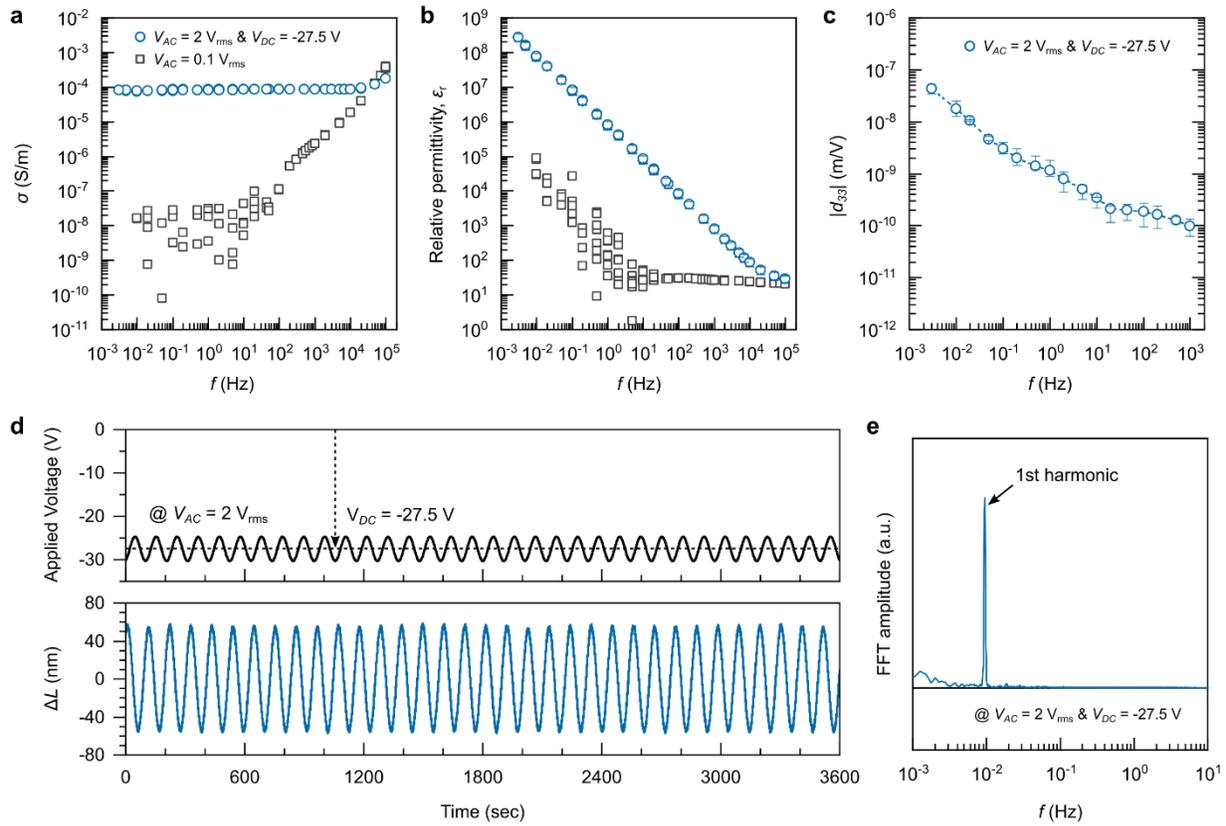

**Fig. S13. Electrical conductivity, dielectric permittivity, and piezoelectricity in YSZ film. (a)** Electrical conductivity versus frequency of a YSZ (Y 8 %) film (~300 nm), measured by AC 100 mV$_{rms}$ (open square) and a simultaneous voltage application (AC 2 V$_{rms}$ and DC -27.5 V). **(b)** Dielectric permittivity versus frequency of the YSZ film. **(c)** Frequency-dependent piezoelectric coefficient, $d_{33}$, of the YSZ film, excited by simultaneous application of AC 2 V$_{rms}$ and DC -27.5 V. **(d)** The 1st-order electromechanical displacements of the YSZ film, excited by simultaneously applying AC 2 V$_{rms}$ and DC -27.5 V ($E_{DC}$ ~0.9 MV/cm), measured at about 10 mHz (9.4 mHz). **(e)** The corresponding FFT magnitude spectra of the output signals in *f*.

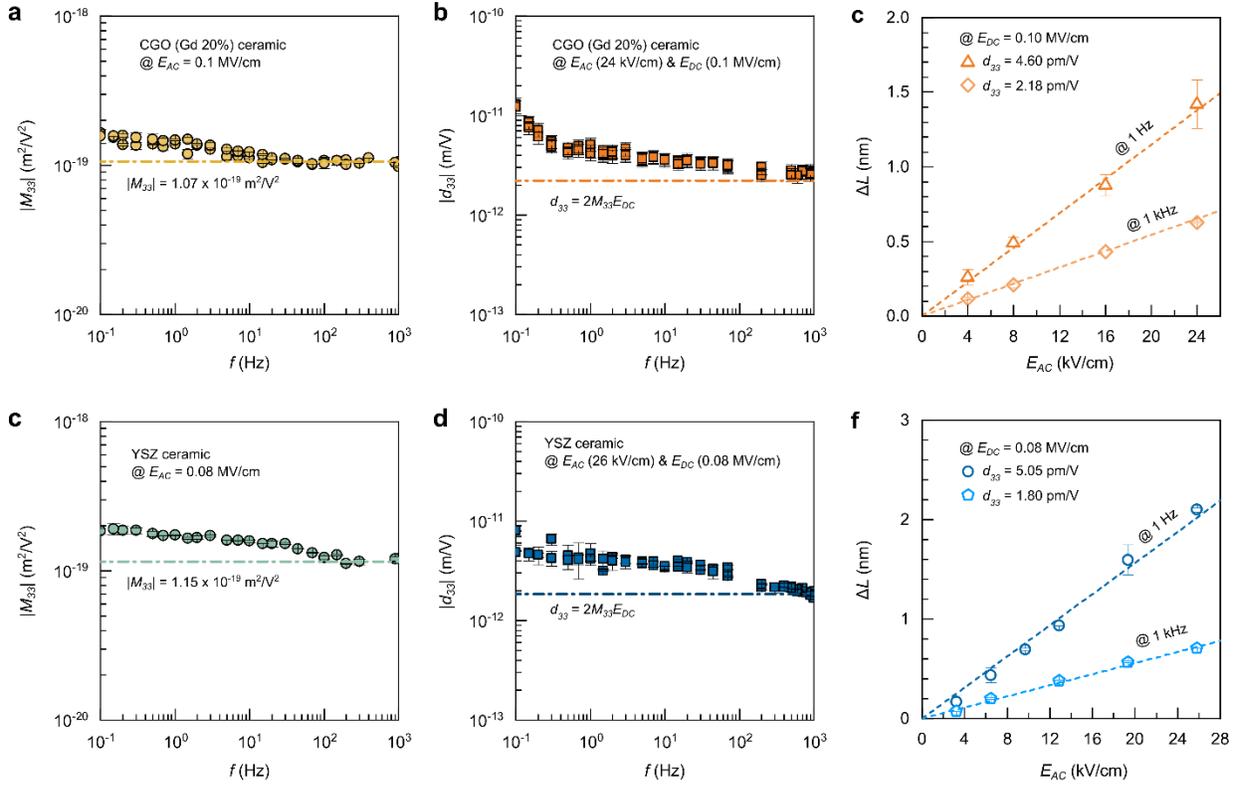

**Fig. S14. Electrostriction and piezoelectricity in CGO and YSZ ceramics.** (**a**) Frequency-dependent $M_{33}$ of CGO ceramic, excited by $E_{AC}$ = 0.1 MV/cm. (**b**) Frequency-dependent $d_{33}$ of the CGO ceramic, excited by $E_{AC}$ (24 kV/cm) + $E_{DC}$ (0.1 MV/cm). (**c**) Linear piezoelectric displacements of the CGO ceramics under $E_{DC}$ = 0.1 MV/cm, measured with various $E_{AC}$ and at 1 Hz and 1 kHz. (**d**) Frequency-dependent $M_{33}$ of YSZ ceramic, excited by $E_{AC}$ = 0.08 MV/cm. (**e**) Frequency-dependent $d_{33}$ of the YSZ ceramic, excited by $E_{AC}$ (26 kV/cm) + $E_{DC}$ (0.08 MV/cm). (**f**) Linear piezoelectric displacements of the YSZ ceramics under $E_{DC}$ = 0.08 MV/cm, measured with various $E_{AC}$ and at 1 Hz and 1 kHz. These results confirm that: (*i*) piezoelectricity in centrosymmetric fluorite oxide ceramics can be generated via our working concept ($E_{AC}$ + $E_{DC}$), and (*ii*) high-frequency (1 kHz) piezoelectricity holds the relation of $d_{33} = 2M_{33}E_{DC}$.

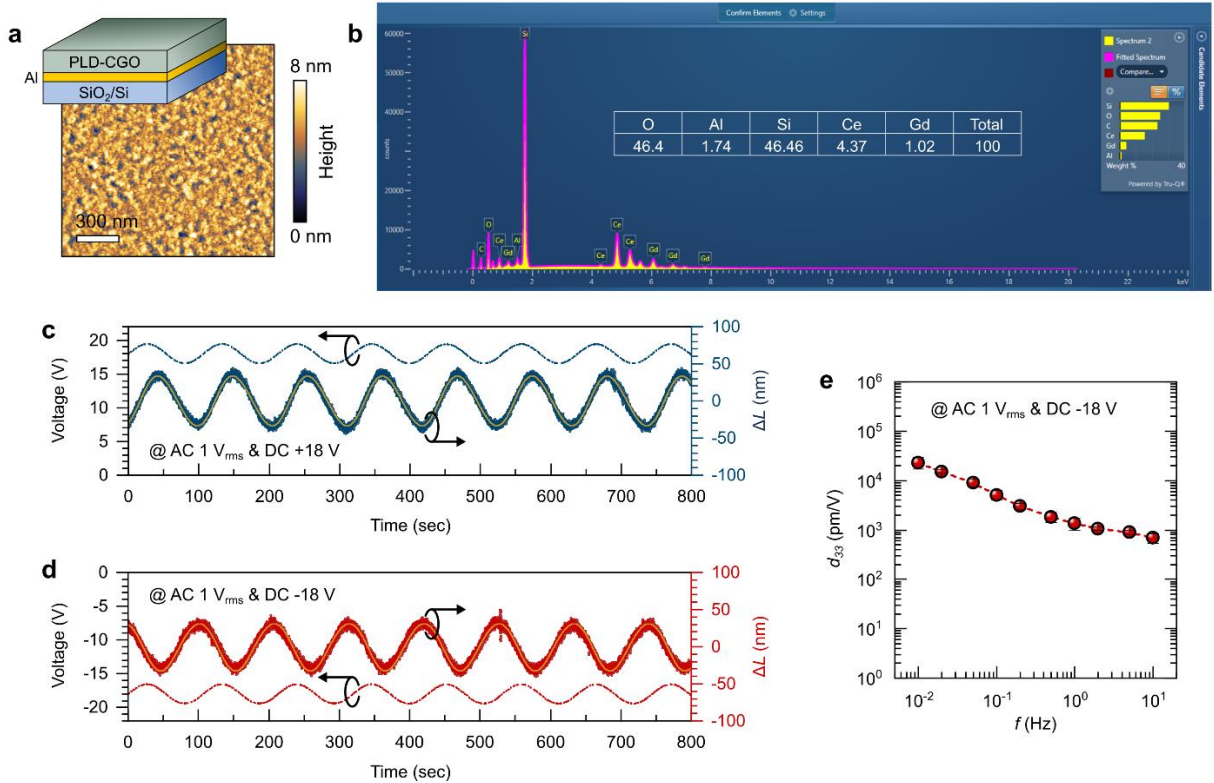

**Fig. S15. Induced piezoelectric effect in a RT-CGO film deposited by using pulsed laser deposition.** **(a)** Surface morphology of a PLD-deposited polycrystalline CGO film (~350 nm) on Al (150 nm)/SiO$_2$ (150 nm)/Si substrate as schematically illustrated. **(b)** Energy dispersive X-ray spectroscopy analysis for the atomic composition of the PLD-deposited CGO sample. No impurities were found in the sample (detection limit <1%). **(c,d)** Reversible switching for the sign of piezoelectric displacements of the PLD-CGO sample with respect to the sign of the applied DC voltages ($E_{DC}$ = ±0.51 MV/cm) while applying the same AC voltage ($E_{AC}$ = 40.4 kV/cm). **(e)** Piezoelectric coefficients ($d_{33}$) of the PLD-deposited CGO film as a function of frequency, $f$. The observed frequency-dependent $d_{33}$ is consistent with that observed in sputtered CGO and PLD-deposited YSZ. Therefore, we confirm that the fundamental mechanism and findings on the generation of the piezoelectricity in CGO are unaffected by Cr impurity and remain the same as discussed in the original paper.

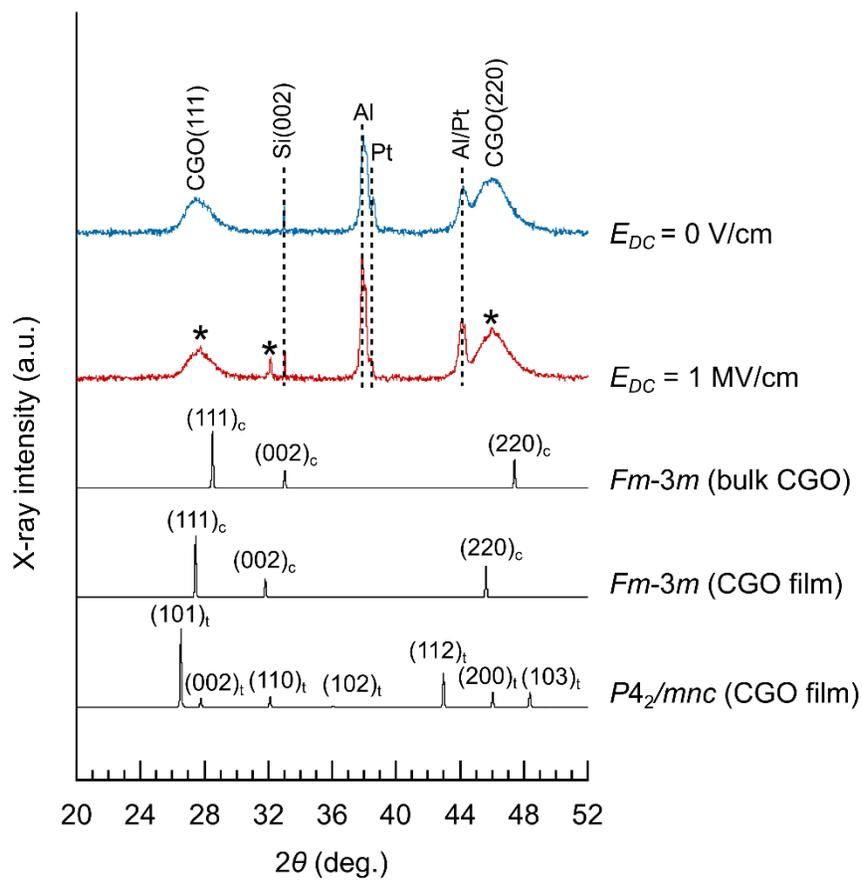

**Fig. S16. *In-situ* X-ray diffraction measurements.** XRD patterns of the CGO film with and without DC electric field application. The pristine polycrystalline CGO film predominantly shows broad (111) and (220) peaks at $2\theta = 27.49°$ and $46.12°$. Whilst, the new peaks visibly appear at $2\theta = 27.78°$, $32.16°$, and $46.03°$ when $E_{DC}$ (= 1 MV/cm) is applied to the CGO film.

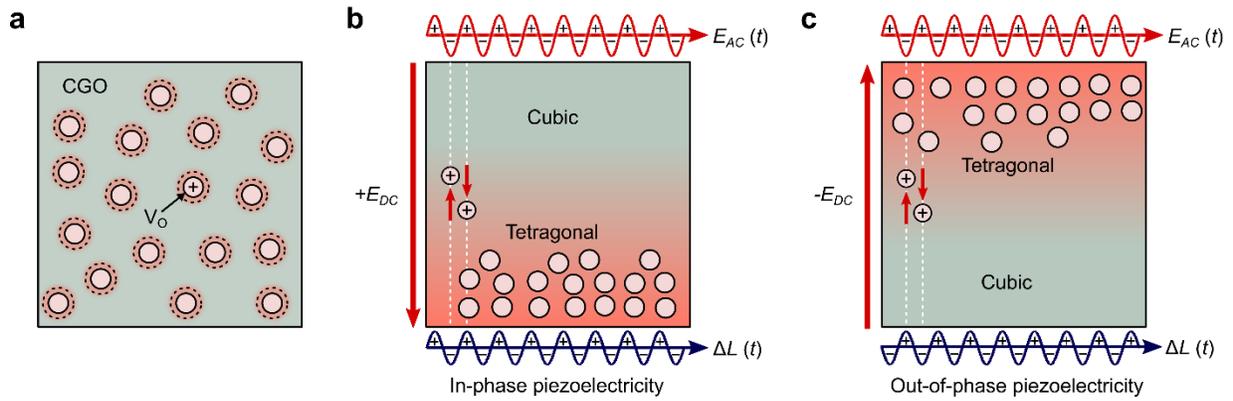

**Fig. S17. Piezoelectric displacements of the CGO film via the control of electric field.** (**a**) The pristine CGO film with a random distribution of oxygen vacancies ($V_O$). $V_O$-induced lattice distortions occur at the local areas of the film. (**b**) In-phase piezoelectric displacements, generated by applying $E_{AC}$ under a concurrent positive $E_{DC}$ (from the top to bottom electrodes). The piezoelectric displacement is determined by $E_{AC}$-driven $V_O$ motion, *i.e.*, more $V_O$ towards the bottom electrode with positive $E_{AC}$ for expansion, while less $V_O$ towards the top electrode with negative $E_{AC}$ for contraction. (**c**) Out-of-phase piezoelectric displacements, generated by applying $E_{AC}$ under a concurrent negative $E_{DC}$ (from the bottom to top electrodes). The piezoelectric displacement is also determined by $E_{AC}$-driven $V_O$ motion, *i.e.*, less $V_O$ towards the bottom electrode with positive $E_{AC}$ for contraction, while more $V_O$ towards the top electrode with negative $E_{AC}$ for expansion.